\documentclass[aps,prb,reprint,superscriptaddress,longbibliography,floatfix]{revtex4-2}

\usepackage{graphicx}
\usepackage{amsmath,amssymb,amsfonts,mathtools}
\usepackage{xcolor}
\colorlet{RED}{red}
\usepackage{hyperref}
\usepackage{booktabs}

\newcommand{\up}{\ensuremath{\uparrow}}
\newcommand{\dn}{\ensuremath{\downarrow}}
\newcommand{\kv}{\mathbf{k}}
\newcommand{\qv}{\mathbf{q}}
\newcommand{\Qv}{\mathbf{Q}}
\newcommand{\Piv}{\textbf{M}}

\newcommand{\id}{\mathbb{I}}

\newcommand{\Se}{\mathcal{S}_e}
\newcommand{\eve}[1]{\ensuremath{\langle #1 \rangle_{\Se}}}

\begin{document}

\title{Investigating competing order in the Kanamori-Hubbard model \\
with the multiorbital multichannel fluctuating field theory}

\author{L. Torchia}
\affiliation{International School for Advanced Studies (SISSA), Via Bonomea 265, 34136 Trieste, Italy}
\author{E. Linn\'er}
\affiliation{International School for Advanced Studies (SISSA), Via Bonomea 265, 34136 Trieste, Italy}
\author{M. Capone}
\affiliation{International School for Advanced Studies (SISSA), Via Bonomea 265, 34136 Trieste, Italy}
\affiliation{CNR-IOM, Istituto Officina dei Materiali, Consiglio Nazionale delle Ricerche, Via Bonomea 265, 34136 Trieste, Italy}

\date{\today}

\begin{abstract}
The multichannel fluctuating field (MCFF) approach treats selected collective fluctuations in interacting quantum lattice models nonperturbatively, but has so far been applied only to single-band systems.
Here, we extend MCFF theory to multiorbital systems, accounting for the orbital structure of charge and spin instabilities.
We apply it to the half-filled two-orbital Kanamori--Hubbard model with repulsive intra-orbital interaction $U$, and both positive and negative Hund's coupling $J$, addressing the competing orders and their interplay.
The ability to selectively retain individual fluctuation channels and control their interplay allows us to disentangle their respective contributions and identify the mechanisms underlying competing instabilities.
The model features a simple competition between charge density wave and antiferromagnetic fluctuations in the vicinity of $J=U$, and a more intricate competition between antiferromagnetic, orbital, and spin-orbital fluctuations near $J=0$, where the model acquires a higher SU(4) symmetry. 
The proposed multiorbital extension thus provides a computationally inexpensive and efficient framework for exploring how orbital and spin character shape competing collective fluctuations.
\end{abstract}

\maketitle

\section{Introduction}
\label{sec:intro}
A striking aspect of correlated materials is the competition between collective instabilities involving charge, spin, pairing and orbital degrees of freedom.
Their interplay can have different outcomes and we can broadly distinguish situations where different orders may suppress one another~\cite{PhysRevB.29.5089,PhysRevLett.53.2327,PhysRevB.108.035143}, or stabilize each other or even lead to additional intermediate phases~\cite{doi:10.1143/JPSJ.68.3123,PhysRevB.61.16377,PhysRevLett.92.236401,PhysRevB.108.205156}.
In multiorbital systems, charge, spin and superconducting order parameters are characterized not only by their spatial and spin structure, but also by their orbital character.
Understanding their interplay requires a framework that retains the orbital content of each fluctuation while allowing competing channels to fluctuate together.
A variety of analytical and computational methods have been introduced and deployed to address these issues, ranging from functional renormalization group~\cite{fRG} to approaches based on the two-particle vertices~\cite{RevDGA} or fully numerical methods such as DMRG or tensor networks~\cite{DMRG_RMP}.
Moreover, mean-field theory~\cite{PhysRevB.83.174513} and dynamical mean-field theory (DMFT)~\cite{PhysRevLett.62.324,Georges1996DMFT,RevModPhys.78.865}, and its cluster~\cite{PhysRevB.89.195146,PhysRevB.79.245128} and diagrammatic extensions~\cite{PhysRevB.95.115107,Galler2018,PhysRevB.103.035120,vanLoon_2021,PhysRevLett.127.207205,SciPostPhys.13.2.036}, have proven insightful in multiorbital models, and \textit{GW}+DMFT~\cite{PhysRevLett.90.086402,Tomczak12,PhysRevLett.109.237010,PhysRevB.88.165119,PhysRevB.88.235110,PhysRevB.90.165138} is routinely applied to realistic multiorbital systems.
Yet, the problem remains prohibitively complex and often computationally heavy and calls for novel approaches to solve it. 

In this paper we introduce, and apply to one of the simplest multi-orbital settings, a multiorbital generalization of the multi-channel fluctuating field method (MCFF)~\cite{PhysRevB.108.035143,PhysRevB.108.205156}.
MCFF is a recently introduced approach that offers an efficient and controlled path to study the interplay of collective instabilities. 
The ancestor of this approach is the  fluctuating (local) field (FF) theory~\cite{PhysRevE.97.052120,PhysRevB.102.224423,PhysRevB.105.035118} developed for magnetic instabilities.
This method improves over static mean-field by replacing a fixed, self-consistent field with the average over an ensemble of collective FFs described by a theory with variationally determined parameters.
As a result, the approach includes the effect of fluctuations beyond mean-field, and allows to compute their free-energy profiles.
MCFF extends the variational construction by incorporating multiple FFs associated with a selection of different leading collective modes~\cite{PhysRevB.108.035143}.
So far the method has been applied to the interplay between charge and magnetic instabilities~\cite{PhysRevB.108.035143} as well as $s$-wave pairing fluctuations~\cite{PhysRevB.108.205156}.
However, these studies focused on the single-orbital Hubbard~\cite{PhysRevB.102.224423,PhysRevB.105.035118} and extended Hubbard models~\cite{PhysRevB.108.035143,PhysRevB.108.205156} with nearest-neighbor interaction, and did not account for any orbital degrees of freedom.
For completeness, besides MCFF theory, we highlight that the original FF theory have been developed in other important directions including strong correlation~\cite{PhysRevB.102.224423,s10948-022-06303-8,PhysRevB.110.245134}, disordered systems~\cite{Kuznetsova2022} and out of equilibrium~\cite{crr8-vwxq}.

Our extension of the method to a multiorbital framework enhances the complexity of the possible solutions.
Moreover, we extend the implementation allowing for an unprecedented number of different FFs, which reveals the potential of the method to introduce significant effects beyond a static mean-field picture.
We notice that this approach improves over static mean-field in a somewhat complementary way with respect to DMFT.
Here an average effect of the fluctuations of the order parameters are included, while DMFT accurately captures local dynamical correlations.

We apply the approach to a minimal yet rich platform for investigating multiorbital physics, the half-filled two-dimensional two-orbital Kanamori-Hubbard model~\cite{Kanamori1963}, which features local inter- and intra-orbital $U$ and $U^\prime$ repulsions, and a Hund's coupling $J$. 
The $J$ coupling plays a key role in distinguishing and favoring different local configurations, with ${J>0}$~\cite{Georges2013,deMedici2011,GeorgesPhysToday,Capone2026} and ${J<0}$~\cite{Capone2009RMP,Capone2orb} promoting high- and low-spin local states, respectively.
At their interface ${J=0}$, for degenerate bands, the model exhibits a SU(4) symmetry emerging from the underlying spin SU(2) and orbital SU(2) symmetries~\cite{Georges2013}.
As Mermin-Wagner-Hohenberg theorem forbids long-range order breaking for continuous symmetry in a large class of two-dimensional systems at finite temperature~\cite{PhysRev.158.383,PhysRevLett.17.1133,PhysRev.171.513,PhysRevLett.107.107201}, including the two-orbital Kanamori-Hubbard model, regions dominated by strong fluctuations associated with these continuous symmetries should not be considered true thermodynamic phases.
Nevertheless, alike in Ref.~\cite{PhysRevB.108.205156}, we will denote such regions of strong fluctuations as ``phases''.
The MCFF construction can also be applied to more general Hamiltonians, including orbital-dependent hopping, crystal-field splittings, and inter-orbital hybridization, although these extensions generally require revisiting the symmetry-driven classification of instabilities that we use in this work.
This flexibility opens the way to studying competing instabilities in more realistic models of multiorbital materials.

Accounting for the interplay between charge and spin channels, while also resolving their orbital structures, we construct a phase diagram at finite temperature as a function of the model parameters, contrasting the regions with positive, zero and negative $J$.
In particular, we first discuss the interplay between different solutions with the charge and the spin channels considered separately, then we move to their joint competition.
We consider, in the latter situation, an unprecedented number of seven simultaneous FFs which are sampled using a Monte Carlo method~\cite{Metropolis1949,Metropolis1953}.
The inclusion of several FFs in competition leads to a region in which neither of the mean-field instabilities is stable, proving a qualitative deviation from mean-field.
In our specific implementation with fixed number of FFs, the region shrinks with the systems size, but we expect that a larger number of FFs can counteract this effect. 
The flexibility to select the fluctuation channels allows us to investigate individual instabilities separately and their mutual interplay.
This ability to decouple each contribution to the joint competition acts as a powerful diagnostic tool to understand the intricate interplay which is present in numerically exact solutions of the model.

The paper is organized as follows:
In Sec.~\ref{sec:model}, we review the model and list the relevant order parameters.
In Sec.~\ref{sec:mcff}, we showcase our proposed multiorbital extension of the MCFF theory, with detailed derivations disclosed in the Supplemental Material, and Sec.~\ref{sec:results} is dedicated to our results from its application to the model.
Finally, Sec.~\ref{sec:conclusions} gives concluding remarks.

\section{Model and order parameters}
\label{sec:model}

To study competing charge, spin and orbital instabilities, we consider a two-orbital Kanamori-Hubbard model at half filling on the square lattice~\cite{Kanamori1963,Werner2007,PhysRevB.79.245128,PhysRevB.81.035112,PhysRevB.83.125110,Werner2orb,Capone2orb,PhysRevLett.129.096404,PhysRevLett.132.226501,Florez,Torchia,Soares2026}, defined by:
\begin{equation}
    \hat H = -t\sum_{\langle i,j\rangle,m,\sigma} \bigl(\hat c^{\dagger}_{im\sigma}\hat c_{jm\sigma}+\mathrm{H.c.}\bigr) + \sum_i \hat H^{(i)}_{\mathrm{int}},
    \label{eq:H}
\end{equation}
with the local Kanamori interactions given by~\cite{Kanamori1963}:
\begin{align}
    \hat H^{(i)}_{\mathrm{int}} = \; & U \sum_{m} \hat n_{im\up} \hat n_{im\dn} + U' \sum_{m\neq m'} \hat n_{im\up} \hat n_{im'\dn}  \nonumber \\
    &+ (U'-J) \sum_{m<m',\sigma} \hat n_{im\sigma} \hat n_{im'\sigma} \nonumber \\
    & - J \sum_{m\neq m'} \hat c^{\dagger}_{im\up} \hat c_{im\dn} \hat c^{\dagger}_{im'\dn} \hat c_{im'\up}  \nonumber \\
    &+ J \sum_{m\neq m'} \hat c^{\dagger}_{im\up} \hat c^{\dagger}_{im\dn} \hat c_{im'\dn} \hat c_{im'\up}.
    \label{eq:Hint}
\end{align}
Here, $\hat c^{(\dagger)}_{im\sigma}$ annihilates (creates) an electron on site $i$ in orbital $m\in\{1,2\}$ with spin $\sigma\in\{\uparrow,\downarrow\}$, and ${\hat n_{im\sigma} \equiv \hat c^{\dagger}_{im\sigma} \hat c_{im\sigma} }$ is its number operator.
The orbitals are degenerate with identical hopping $t$ between nearest-neighbors $\langle i,j\rangle$, without any inter-orbital hopping.
We set $t \equiv 1$ as the unit of energy.
The parameters $U$ and $U'$ describe the intra-orbital and inter-orbital density interactions, respectively, and Hund's coupling $J$ enters in the spin-flip and pair-hopping terms.
We adopt the rotationally invariant relation $U' \equiv U-2J$~\cite{Castellani1978}.

In this work, we consider a repulsive intra-orbital interaction $U>0$, and both repulsive and attractive regimes of Hund's coupling $J$.
For real materials, the conventional physical regime is $0<J\lesssim U/3$, where both inter-orbital density-interaction coefficients, $U'=U-2J$ and $U'-J=U-3J$, are positive, corresponding to purely repulsive interactions~\cite{Castellani1978,Georges2013}.
Nevertheless, negative values of $J$, corresponding to an inverted Hund's coupling, can also arise in physical systems, for instance through Jahn--Teller phonons~\cite{Capone2009RMP}.
For $J>U/3$, the parallel-spin inter-orbital interaction $U'-J$ becomes attractive, followed by the opposite-spin inter-orbital interaction $U'$ for $J>U/2$.
To characterize these competing charge, spin and orbital instabilities, we introduce a large selection of order parameters in the particle-hole channel, which we are considering in this work.

\subsection{Order parameters}
\label{sec:instabilities_ph}
In this section we introduce all the different orders that we address in this work.
In principle, we can describe all the possible particle-hole instabilities accessible to the MCFF starting from the sixteen different local particle-hole operators that can be defined in our two-orbital system, which form a complete basis for the local particle-hole bilinears:
\begin{equation}
    \begin{aligned}
    \hat n_j^{\varrho,\varsigma} & \equiv \sum_{\{m\},\{\sigma\}} \hat c^{\dagger}_{jm\sigma}\, \tau^{\varrho}_{mm'}\sigma^{\varsigma}_{\sigma\sigma'}\, \hat c_{jm'\sigma'}.
    \end{aligned}
    \label{eq:local_composites}
\end{equation}
Here $\tau^\varrho$ and $\sigma^\varsigma$ are matrices in the orbital and spin space, respectively, with $\varrho\in\{0,x,y,z\}$ and $\varsigma\in\{c,x,y,z\}$ spanning the identity matrix and the three Pauli matrices.
In the second index we used the symbol $c$ to highlight that the unit matrix corresponds to the sum over the two spins, i.e., the total occupation or charge on the site/orbital, while the Pauli matrices naturally describe the three components of the electron spin. 
We notice that an analogous construction can be introduced for models with $N_{\mathrm{orb}}>2$ orbitals, where the orbital Pauli matrices are replaced by the identity and the generators of $\mathrm{SU}(N_{\mathrm{orb}})$.
For example, for $N_{\mathrm{orb}}=3$, these generators are the eight Gell-Mann matrices.

\begin{figure}[t!]
\centering
\includegraphics[width=0.92\columnwidth]{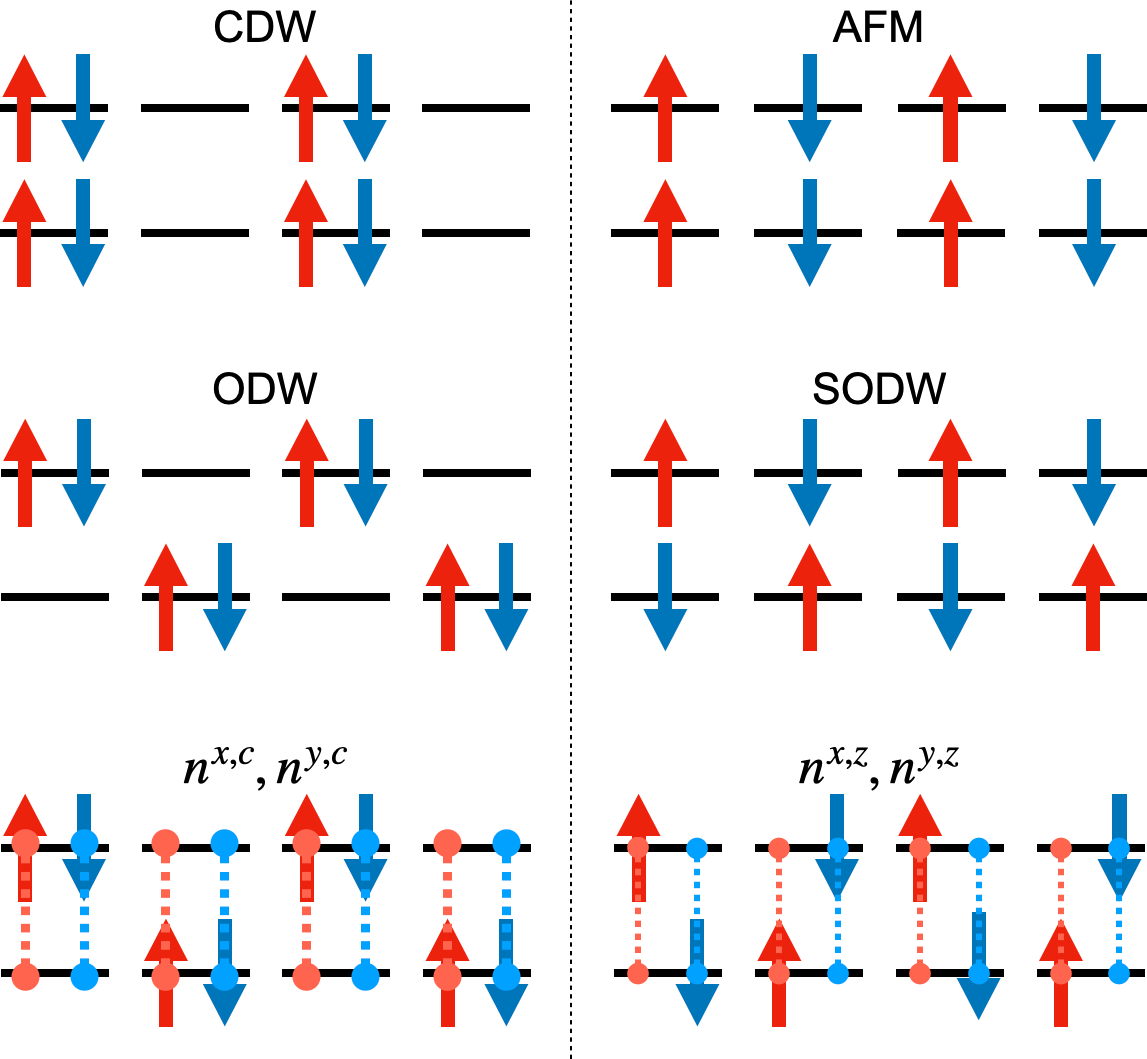}
\caption{Illustration of the particle-hole instabilities in the two-orbital Kanamori-Hubbard model addressed in this work.}
\label{fig:order}
\end{figure}

For any of the bilinears in Eq. (\ref{eq:local_composites}), we can have different wave vectors $\Qv$ which specify the spatial pattern $e^{i \Qv \cdot {\bf r}}$, with the position ${\bf r}=(j_x,j_y)$.
In this work, where we are limited to half filling, we can focus on staggered phases with characteristic vector $\Qv=\Piv \equiv(\pi,\pi)$.
This is due to the perfect nesting of the Fermi surface following from the bare dispersion ${\epsilon_\kv=-2(\cos k_x+\cos k_y)}$ satisfying $\epsilon_{\kv+\Piv}=-\epsilon_\kv$, where $\kv$ is the crystal momentum.
We remain however with several possible competing modes, that we highlight here. 

We now provide a short description of the relevant modes.
In particular we classify some of these states using more transparent names in order to simplify the discussion.
In the remainder of this section we omit the site subscript $i$ and a staggered $(-1)^{j}$ factor to keep the notation as light as possible. 
In all the situations where the spin part is not the unit matrix, we will only report the case of the magnetization along the z-axis for simplicity. 

The two orbital-homogeneous modes, associated with the $\tau^{0}$-matrix, correspond to the checkerboard CDW and N\'eel AFM, which are defined by:
\begin{align}
   \text{CDW}: n^{0,c}& \propto \sum_{m=\{1,2\}}[ \hat{n}_{m\uparrow} + \hat{n}_{m\downarrow}],\\
    \text{AFM}: n^{0,z}& \propto \sum_{m=\{1,2\}}[ \hat{n}_{m\uparrow} - \hat{n}_{m\downarrow}].
\end{align}
On the contrary, the channels associated with the $\tau^{z}$-matrix are connected to phases with an orbital polarization.
We define orbital density wave (ODW) and spin-orbital density wave (SODW), with its spin component being uniform or staggered, respectively:
\begin{align}
    \text{ODW}: n^{z,c} & \propto  \sum_{\sigma=\{\uparrow,\downarrow\}}[ \hat{n}_{1\sigma} - \hat{n}_{2\sigma}],\\
    \text{SODW}: n^{z,z} & \propto  \hat{n}_{1\uparrow} - \hat{n}_{1\downarrow} - \hat{n}_{2\uparrow} + \hat{n}_{2\downarrow}.
\end{align}
The remaining instabilities, associated with orbital matrices $\tau^x$ or $\tau^y$ are associated with hybridization between different orbitals, connected with excitonic order.
We avoid to introduce an extra set of specific names for these phases since they play a minor role in our analysis.
These remaining independent order parameters are thus: 
\begin{align}
n^{x,c} &\propto  \hat{c}^{\dagger}_{
1\uparrow} \hat{c}_{2\uparrow} + \hat{c}^{\dagger}_{
2\uparrow} \hat{c}_{1\uparrow} + \hat{c}^{\dagger}_{1\downarrow} \hat{c}_{2\downarrow} + \hat{c}^{\dagger}_{2\downarrow} \hat{c}_{1\downarrow} ,\\
n^{y,c} & \propto  
        \hat{c}^{\dagger}_{1\uparrow} \hat{c}_{2\uparrow} - \hat{c}^{\dagger}_{2\uparrow} \hat{c}_{1\uparrow} + \hat{c}^{\dagger}_{1\downarrow} \hat{c}_{2\downarrow} - \hat{c}^{\dagger}_{2\downarrow} \hat{c}_{1\downarrow} , \\
n^{x,z} & \propto \hat{c}^{\dagger}_{1\uparrow} \hat{c}_{2\uparrow} + \hat{c}^{\dagger}_{2\uparrow} \hat{c}_{1\uparrow} - \hat{c}^{\dagger}_{1\downarrow} \hat{c}_{2\downarrow} - \hat{c}^{\dagger}_{2\downarrow} \hat{c}_{1\downarrow},\\
n^{y,z} & \propto
    \hat{c}^{\dagger}_{1\uparrow} \hat{c}_{2\uparrow} - \hat{c}^{\dagger}_{2\uparrow} \hat{c}_{1\uparrow} - \hat{c}^{\dagger}_{1\downarrow} \hat{c}_{2\downarrow} + \hat{c}^{\dagger}_{2\downarrow} \hat{c}_{1\downarrow}.
\end{align}
Throughout this work, the ODW and SODW modes are degenerate with the orbital-hybridization modes $n^{x,c}$ and $n^{x,z}$, respectively.
Our calculations account for both components of each degenerate pair through their combined radial amplitude.
However, for notational simplicity, we refer to these instabilities by their orbital-polarized representatives, ODW and SODW.

\subsection{Action formalism}
\label{sec:KH_Action}
We introduce the MCFF theory using a path integral action formalism. 
The action of the considered two-orbital Kanamori-Hubbard model is given by:
\begin{align}
    \mathcal{S} = & - \frac{1}{\beta N_s}\sum_{k,m,\sigma} c^{*}_{k m\sigma}\, \mathcal{G}^{-1}_{k}\, c_{k m\sigma}  + \mathcal{S}_{\text{int}},
    \label{eq:S}
\end{align}
with the number of sites $N_s=L^2$ on a $L\times L$ lattice, and the inverse temperature $\beta = 1/k_B T$. 
The Grassmann variables $c^{(*)}_{k m\sigma}$ describe the annihilation (creation) of an electron, with the subscripts $k=(\nu,\kv)$ combining the fermionic Matsubara frequency $\nu$ and crystal momentum $\kv$.
The bare inverse propagator $\mathcal G_k^{-1}=i\nu+\mu-\epsilon_\kv$ is identical for the two orbitals, and depends on the dispersion ${\epsilon_\kv}$ and chemical potential $\mu$.
Finally, the local interactions of the action is easily obtained from the Hamiltonian:
\begin{align} 
    \mathcal{S}_{\text{int}} = & U \sum_{m} \rho_{q,m,\uparrow} \rho_{-q,m,\downarrow}
     + U^\prime \sum_{m\neq m^\prime} \rho_{q,m,\uparrow} \rho_{-q,m^\prime,\downarrow} \nonumber \\
     & + (U^\prime-J) \sum_{m < m^\prime, \sigma} \rho_{q,m,\sigma} \rho_{-q,m^\prime,\sigma} \nonumber \\
    & -J \sum_{m\neq m^\prime,k} c^{*}_{k+q,m,\uparrow} c_{k,m,\downarrow} c^{*}_{k^\prime-q,m^\prime,\downarrow} c_{k^\prime,m^\prime, \uparrow}  \nonumber \\
    & + J \sum_{m\neq m^\prime,k} c^{*}_{k+q,m,\uparrow}c^{*}_{k^\prime-q,m,\downarrow} c_{m^\prime,k,\downarrow} c_{m^\prime,k^\prime,\uparrow}, \label{Eq:S_int}
\end{align}
with the shifted densities ${\rho_{q,\sigma,m} = n_{q,\sigma,m} - \langle n_{q,\sigma,m}\rangle\delta_{q,0}}$ with ${n_{q,\sigma,m} = \sum_{k} c^{*}_{k+q,\sigma,m} c^{\phantom{*}}_{k,\sigma,m}}$.
Here the subscripts ${q=(\omega,\qv)}$ combine the bosonic Matsubara frequency $\omega$ and crystal momentum $\qv$.

\section{Multiorbital multichannel fluctuating field theory}
\label{sec:mcff}

We generalize the MCFF theory, recently developed to describe competing collective instabilities in the single-orbital extended Hubbard model~\cite{PhysRevB.108.035143,PhysRevB.108.205156}, from its single-orbital origin to a multiorbital construction.
Thus, we aim to permit for a general description of orbital-resolved collective instabilities in the particle-hole channels, and their interplay.
In the spirit of the prior single-orbital formulations of the FF theory~\cite{PhysRevE.97.052120,PhysRevB.102.224423,PhysRevB.105.035118,s10948-022-06303-8,PhysRevB.108.035143,PhysRevB.108.205156}, we introduce the multiorbital FF action:
\begin{align}
    \mathcal{S}^{*} = & - \frac{1}{\beta N_s}\sum_{k,m,\sigma} c^{*}_{k m\sigma}\, \mathcal{G}^{-1}_{k}\, c_{k m\sigma} \nonumber \\
    & + \sum_{\alpha,\Qv} \left[ \phi^{\alpha}_{\Qv}\, \rho^{\alpha}_{-\Qv} - \frac{1}{2} \frac{\beta N_s}{J^{\alpha}_{\Qv}}\, \phi^{\alpha}_{\Qv} \phi^{\alpha}_{-\Qv} \right],
    \label{eq:trial}
\end{align}
restricted to a set of retained FFs $\phi^{\alpha}_{\Qv}$.
Each FF $\phi^{\alpha}_{\Qv}$ is coupled to a static composite variable ${\rho^{\alpha}_{\Qv}=n^{\alpha}_{\Qv}-\langle n^{\alpha}_{\Qv}\rangle\delta_{\Qv,0}}$ corresponding to the $\alpha$ channels that are retained.
Limiting ourselves to the particle-hole channels, the composite variables are given by $n_{\Qv}^{\varrho,\varsigma}$, which denotes the corresponding Grassmann bilinear, with $\alpha=(\varrho,\varsigma)$ labeling its spin and orbital structure.
For each FF $\phi^{\alpha}_{\Qv}$, we compute a stiffness parameter $J^{\alpha}_{\Qv}$, via the Peierls--Feynman--Bogoliubov variational prescription~\cite{PhysRev.54.918,Bogolyubov:1958zv,feynman1972}.
This is based on minimizing a variational free-energy functional relative the original action, as performed in the single-orbital FF approaches~\cite{PhysRevB.102.224423,PhysRevB.108.035143,PhysRevB.108.205156}, fixing the stiffness parameters without any Fierz decoupling~\cite{PhysRevD.68.025020,PhysRevB.70.125111,jaeckel2002understanding} (SM Secs.~S1 and S2).
Furthermore, our choice of channel basis makes the resulting quadratic interaction functional diagonal.

For each FF configuration, the fermions can be integrated out explicitly, yielding an effective action for the FFs degrees of freedom:
\begin{equation}
    \mathcal{S}_{\phi} = - \mathrm{Tr}\ln\Big[ \mathcal{G}^{-1} - \sum_{\alpha,\Qv} \phi^{\alpha}_{\Qv} \Lambda^{\alpha}_{-\Qv} \Big] - \frac{1}{2}\sum_{\alpha,\Qv} \frac{\beta N_s}{J^{\alpha}_{\Qv}}\, \phi^{\alpha}_{\Qv} \phi^{\alpha}_{-\Qv},
    \label{eq:Sphi}
\end{equation}
with the matrix $\Lambda^{\alpha} \equiv \tau^{\varrho}\sigma^{\varsigma}$, from the definition of $n^{\alpha}$.
This action defines a non-quadratic effective free energy, and integrating over the selected FFs then captures their thermal fluctuations beyond the Gaussian approximation. 
With our sign convention, the real-FF integral converges only for $J^\alpha_{\Qv}<0$, and therefore only such channels are retained.
Different levels of approximation can be achieved retaining different FF modes, with the limit of all ordering vectors $\Qv$ and frequency modes $\omega$ included leading to the recovery of the exact description~\cite{PhysRevB.108.035143}.

Integrating over all FFs except for one, allows to define the free energy profile for the one remaining FF:
\begin{equation}
    \mathcal F_\alpha(\phi^\alpha) = -\frac{1}{\beta}\ln\int\prod_{\alpha'\neq\alpha}d\phi^{\alpha'}\,e^{-\mathcal S_\phi}.
    \label{eq:F_alpha}
\end{equation}
Following Refs.~\cite{PhysRevB.108.035143,PhysRevB.108.205156}, a global minimum of $\mathcal F_\alpha$ for a non-zero value of the FF implies that the channel spontaneously develops a finite amplitude, while a secondary minimum identifies a metastable state arising due to the interplay with the integrated FFs.
For few retained modes, the numerical integration in Eq.~\ref{eq:F_alpha} may be performed using the trapezoidal rule over a sufficiently dense grid.
In Sec.~\ref{sec:charge_vs_spin}, we present results for the joint charge--spin competition in the two-orbital model which involves seven FFs simultaneously.
For this unprecedented number of FFs, we instead use Markov chain Monte Carlo for the numerical integration~\cite{Metropolis1949,Metropolis1953}.
The explicit free energy and integration measures are given in SM Secs.~S3 and S4.

\subsection{Stiffness parameters}
\label{sec:composites}
Using the Peierls--Feynman--Bogoliubov variational principle, we obtain the stiffness parameters determining the MCFF action $\mathcal{S}^{*}$ (SM Table~S1):
\begin{align}
    J^{0,c} & = \tfrac14(U+2U'-J), & J^{0,s} & = -\tfrac14(U+J), \nonumber\\
    J^{x,c} & = \tfrac14(-U'+3J), & J^{x,s} & = -\tfrac14(U'+J), \nonumber\\
    J^{y,c} & = \tfrac14(-U'+J), & J^{y,s} & = \tfrac14(-U'+J), \nonumber\\
    J^{z,c} & = \tfrac14(U-2U'+J), & J^{z,s} & = \tfrac14(-U+J),
    \label{eq:J_ph}
\end{align}
where it follows from the spin SU(2) symmetry that all spin components $s \in \{ x,y,z \}$ are degenerate. 
For $U' \equiv U-2J$, the values are collected in Table~\ref{tab:stiffness} for the different phases.

Comparing the values of the different stiffness parameters can help to determine the characteristic energy scales associated with the competition between the different phases, and permit drawing a tentative phase diagram.
We can start with two simple observations, setting aside a more detail accounting for later. 
(i) For the rotationally invariant choice ${U' \equiv U-2J}$: ${J^{x,c}=J^{z,c}}$ and ${J^{x,s}=J^{z,s}}$, which reflects the orbital rotation generated by $\tau^y$.
For a generic value of  $U'$, the two doublets split with ${J^{x,c}-J^{z,c}=J^{z,s}-J^{x,s}=(U'-U+2J)/4}$.
(ii) At $J=0$ all the channels except the CDW instability have the same stiffness $J^{\varrho,\varsigma}=-U/4$, with $(\varrho,\varsigma)\neq(0,c)$, reflecting their degeneracy.
This emerges since, for $J=0$, the model Hamiltonian $\hat H_{\mathrm{int}}=\tfrac U2\hat N(\hat N-1)$, with the on-site occupation $\hat N$, is SU(4) symmetric.
The enhanced symmetry at $J=0$ then provides an organizing point for the competing channels.
This point is analogous to the pseudospin-symmetric line which emerges in the attractive extended Hubbard model~\cite{PhysRevB.108.205156}, although here the degenerate channels are AFM, ODW, SODW, and the orbital-hybridization modes, rather than the CDW and ${s\text{-wave}}$ superconductivity.

\begin{table}[!tbp]
\caption{Correspondence between the indices of Eq.~\eqref{eq:local_composites}, phase names, and stiffness parameters for $U'\equiv U-2J$.
Spin-sector entries refer to the collinear component $\varsigma=z$.
FFs are retained only for $J^{\varrho,\varsigma}<0$, with the single-channel onset given by $J^{\varrho,\varsigma}<-U_c/4$ (SM Sec.~S3).}
\label{tab:stiffness}
\begin{ruledtabular}
\begin{tabular}{llll}
$(\varrho,\varsigma)$ & Phase & Stiffness & Value \\
\hline
\multicolumn{4}{l}{\textit{Charge sector}} \\
$(0,c)$ & CDW & $J^{0,c}$ & $\tfrac14(3U-5J)$ \\
$(z,c)$ & ODW & $J^{z,c}$ & $\tfrac14(-U+5J)$ \\
$(x,c)$ & $n^{x,c}$ & $J^{x,c}$ & $\tfrac14(-U+5J)$ \\
$(y,c)$ & $n^{y,c}$ & $J^{y,c}$ & $\tfrac14(-U+3J)$ \\
\hline
\multicolumn{4}{l}{\textit{Spin sector}} \\
$(0,z)$ & AFM & $J^{0,z}$ & $-\tfrac14(U+J)$ \\
$(z,z)$ & SODW & $J^{z,z}$ & $\tfrac14(-U+J)$ \\
$(x,z)$ & $n^{x,z}$ & $J^{x,z}$ & $\tfrac14(-U+J)$ \\
$(y,z)$ & $n^{y,z}$ & $J^{y,z}$ & $\tfrac14(-U+3J)$ \\
\end{tabular}
\end{ruledtabular}
\end{table}

\subsection{General Considerations on Competing Orders}
\label{sec:spectrum}
In this subsection we discuss some fundamental relations connecting mathematical properties of the different FFs and their interplay. 
For static FFs with ordering $\Qv = \Piv$, we can cast the quadratic Hamiltonian ${H_\kv \equiv \epsilon_\kv +\sum_{\alpha,\Qv} \phi^{\alpha}_{\Qv} \Lambda^{\alpha}_{-\Qv}}$, describing the internal fermionic contribution to the effective FF action $\mathcal{S}_{\phi}$~\eqref{eq:Sphi}, in the basis of eight-component spinors $(c_\kv,c_{\kv+\Piv})$ as
\begin{equation}
 H_\kv=\begin{pmatrix}\epsilon_\kv\id_4&\Phi\\
                    \Phi&-\epsilon_\kv\id_4\end{pmatrix},\hspace*{0.2cm}\text{with}\hspace*{0.2cm}
 \Phi=\sum_\alpha\phi^\alpha\,\Gamma^\alpha.
 \label{eq:Hk}
\end{equation}
Here $\Gamma^\alpha=\tau^{\varrho}\otimes\sigma^{\varsigma}$.
Since the kinetic term anticommutes with the term containing the FFs, the eigenvalues of the Hamiltonian $H_\kv$ are $\pm\sqrt{\epsilon_\kv^2+\lambda^2}$, where $\lambda$ are the eigenvalues of $\Phi$. Hence $\lambda$ act as spectral gaps for the fermions.
Every ${\Gamma^\alpha}$ squares to the identity, so every single FF gives ${\lambda^2=(\phi^\alpha)^2}$.
Therefore, all channels share the same onset ${J^\alpha=-U_c/4}$, and at the saddle-point level the stiffness hierarchy of Table~\ref{tab:stiffness} alone decides which channel prevails.

When we consider two competing FFs, we can draw very general conclusions based on the algebra of their matrices. 
If $\Gamma^1$ and $\Gamma^2$ anticommute, ${\Phi^2=(\phi_1^2+\phi_2^2)\id_4}$, then the FFs add in quadrature in the eigenvalues: ${\lambda^2=\phi_1^2+\phi_2^2}$.
It follows that the fermionic contribution only depends on their radius, with the quadratic stiffness term selecting the orientation.
Moreover, at equal stiffness terms the rotation between the two channels is flat, involving zero free-energy cost.
Examples in the charge sector ($\varsigma=c$) are ${(\phi^{x,c},\phi^{y,c},\phi^{z,c})}$, or in the collinear-spin sector ($\varsigma=z$) are ${(\phi^{x,z},\phi^{y,z},\phi^{z,z})}$, which at $J=0$ admits continuous rotation at fixed orbital-vector length.
Whenever the full spectrum depends on a group of equal-stiffness FFs only through their combined radius, their angular variables can be integrated out, retaining the radial Jacobian in the FF measure.
We use this reduction for the doublets $(\phi^{x,c},\phi^{z,c})$ and $(\phi^{x,z},\phi^{z,z})$ in the separate charge and spin sector calculations, respectively, with measure $2\pi R\,\mathrm{d}R$ (SM Sec.~S4).

If $\Gamma^1$ and $\Gamma^2$ commute, being trace-orthogonal, the FFs add according to $|\phi_1|\pm|\phi_2|$, with one FF allowed to close as the competing amplitude grows, such that separated minima and metastability may appear.
An example is the competing CDW and AFM, with an analogous structure in the single-orbital MCFF~\cite{PhysRevB.108.035143}.
Thus, the matrix algebra determines how competing FFs combine in the spectrum, but establishing whether their competition produces a first- or second-order transition requires the full free energy.
Nevertheless, it gives a particularly transparent distinction between the orbital-homogeneous and orbital-inhomogeneous channels.
For example, whereas the charge-sector channels ${(\phi^{x,c},\phi^{y,c},\phi^{z,c})}$ mutually anticommute, they all commute with the CDW mode $\phi^{0,c}$.
Likewise, an identical distinction holds for the collinear spin-sector channels.

\section{Channel hierarchy and competing fluctuations in the two-orbital Hubbard model}
\label{sec:results}

In this section we move to the analysis of the two-orbital Kanamori-Hubbard model  $L\times L$ lattices with periodic boundary conditions, with ${L\in\{4,8,16,32\}}$.
We set the temperature at $\beta=10/t$, convenient to compare with earlier work, and highlight that MCFF method is applicable for both higher and lower temperatures.
Within our generalize multiorbital MCFF theory, we consider the leading charge, spin and orbital FFs with momentum $\Qv=\Piv$ introduced above.
We first establish the hierarchy between the different collective instabilities at the saddle-point level, which amounts to minimizing the joint free energy of all FFs after integrating out the fermions, but without integrating over any of the auxiliary FFs, i.e. without accounting for the explicit competition between the different modes.
Then, we examine how the integration over the competing FFs change their behavior near their degeneracy lines, where their interplay plays an important role. 

\subsection{Hierarchy of single-channel instabilities and onsets}
\label{sec:hierarchy}
Due to the identical and nested dispersions, a single normalized FF produces the same fermionic spectrum in all channels considered.
Moreover, the free energy $\mathcal{F}_{\alpha}(\phi^{\alpha})$ of the single-channel FF allows for an analytic identification of its phase boundary, which is given by the critical value of $J^\alpha$ for which the curvature of $\mathcal{F}_{\alpha}(\phi^{\alpha})$ changes sign at $\phi^{\alpha}=0$.
We observe the sign change to occur at $J^\alpha=-U_c(L)/4$ (SM Sec.~S3), with
\begin{equation}
 U_c(L)=\left[\frac{1}{N_s}\sum_{\kv\in\mathrm{BZ}}
 \frac{\tanh(\beta\epsilon_\kv/2)}{2\epsilon_\kv}\right]^{-1},
 \label{eq:onset_main}
\end{equation}
being satisfied for the ordering $\Qv = \Piv$.
Eq.~\eqref{eq:onset_main} is the finite-lattice form of the linearized Hartree--Fock AFM gap condition~\cite{LangmannLenells2025}, expressed in our channel normalization (SM Sec.~S3).
For $\beta=10/t$, it yields ${U_c(32)\simeq1.4743}$ and ${U_c(\infty)\simeq1.4775}$; we use the latter as the reference throughout our current overview.

\begin{figure}[t!]
\centering
\includegraphics[width=0.90\columnwidth]{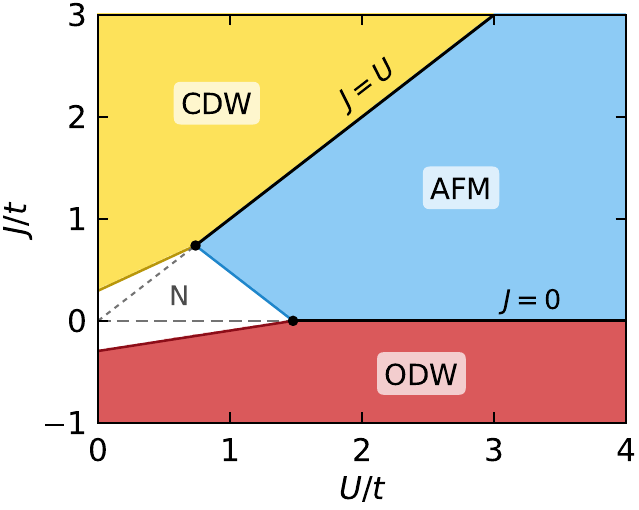}
\caption{Analytical saddle-point hierarchy for $U'\equiv U-2J$ at $\beta=10/t$, using $U_c(\infty)$.
Colored regions indicate ODW (red), AFM (blue), and CDW (yellow); white denotes the normal state.
The ODW mode $n^{z,c}$ is degenerate with the orbital-hybridization mode $n^{x,c}$.
Colored lines mark onset thresholds; black lines mark equal stiffness parameters, solid within the ordered regions and dashed or dotted elsewhere.
Dots mark $(U_c,0)$ and $(U_c/2,U_c/2)$.}
\label{fig:hierarchy}
\end{figure}

We consider the leading particle--hole channels for ${U>0}$, with their associated stiffness parameters $J_{\bf Q}^{\alpha}$ listed in Table~\ref{tab:stiffness}.
Within the single-channel saddle-point treatment, a phase diagram can be drawn simply identifying the channels with the most negative stiffness parameter for every value of the parameters.
Hence the ODW is favored for ${J<0}$, whereas orbital-homogeneous AFM and CDW are favored for ${0<J<U}$ and ${J>U}$, respectively.
For each of these channels, we can define a (single-channel) onset condition
\begin{equation}
\begin{aligned}
  U-5J&>U_c,  &&J<0,\\
   U+J&>U_c,   &&0<J<U,\\
 5J-3U&>U_c, &&J>U.
\end{aligned}
\label{eq:leading_onsets}
\end{equation}
These onset conditions and stiffness crossings provide a simple analytical characterization of the phase diagram that we report in Fig.~\ref{fig:hierarchy}, with ODW (red), AFM (blue), and CDW (yellow) phases. 
The diagram reflects the natural tendencies driven by the various couplings: repulsive $U$ promotes AFM and ODW instabilities but suppresses CDW, while positive $J$ favors AFM and CDW, whereas negative $J$ drives ODW due to the energetic advantage to unbalance the occupation of the two orbitals.

The leading channel changes when we cross the two equal-stiffness lines: the ODW--AFM boundary at $J=0$ for $U>U_c$, and the AFM--CDW boundary at $J=U$ for $U>U_c/2$, shown as solid black lines in Fig.~\ref{fig:hierarchy}.
At $J=0$, we highlight that the model has an exact SU(4) symmetry (see Sec.~\ref{sec:composites}), which makes all fifteen traceless particle--hole channels degenerate, with stiffness $-U/4$.
At the transition line $J=U$, the AFM and CDW stiffness parameters coincide at $-U/2$, giving the common onset at ${U=J=U_c/2}$.
Along these two transition lines, degenerate finite-amplitude minima occur at the saddle-point level for ${U>U_c}$ at $J=0$ and ${U>U_c/2}$ at $J=U$.
Note, the saddle-point value $U_c$ for the AFM phase boundary at $J=0$ is in exact agreement with its equivalent MCFF theory estimate for the analogous half-filled single-orbital Hubbard model at $\beta=10/t$~\cite{PhysRevB.108.035143}.

These general trends are in agreement with observations in Ref.~\cite{Werner2orb} for the multiorbital DMFT applied on the low-temperature regime of the two-orbital model on the Bethe lattice.
In particular, they observed $J$ driving the AFM and CDW phases for $J>0$, and ODW for $J<0$.
However, along with these observations, they capture pairing instabilities and correlation-driven phases, such as, e.g., conventional and paired Mott insulators, which are not accounted for within our method.
Nevertheless, these neglected phases only emerge at sufficiently large $\vert J \vert$, whereas our work will be mainly dedicated to the small $J$ regime, predicted by multiorbital DMFT to showcase leading AFM and ODW fluctuations~\cite{Werner2orb}.

In the following, we focus on the region around $J=0$, where orbital degrees of freedom play a central role.
We highlight that the AFM--CDW competition near $J=U$ has in fact already been studied in an analogous situation in the half-filled extended Hubbard model~\cite{PhysRevB.108.035143}.
Since both orders are orbital-homogeneous, we do not expect the additional orbital degree of freedom to drastically change their competition qualitatively.
In the next sections we move beyond the saddle-point limit, to consider the competition of different modes.

\subsection{Competition in the charge sector}
\label{sec:res_charge}
\begin{figure}[b!]
\centering
\includegraphics[width=0.94\columnwidth]{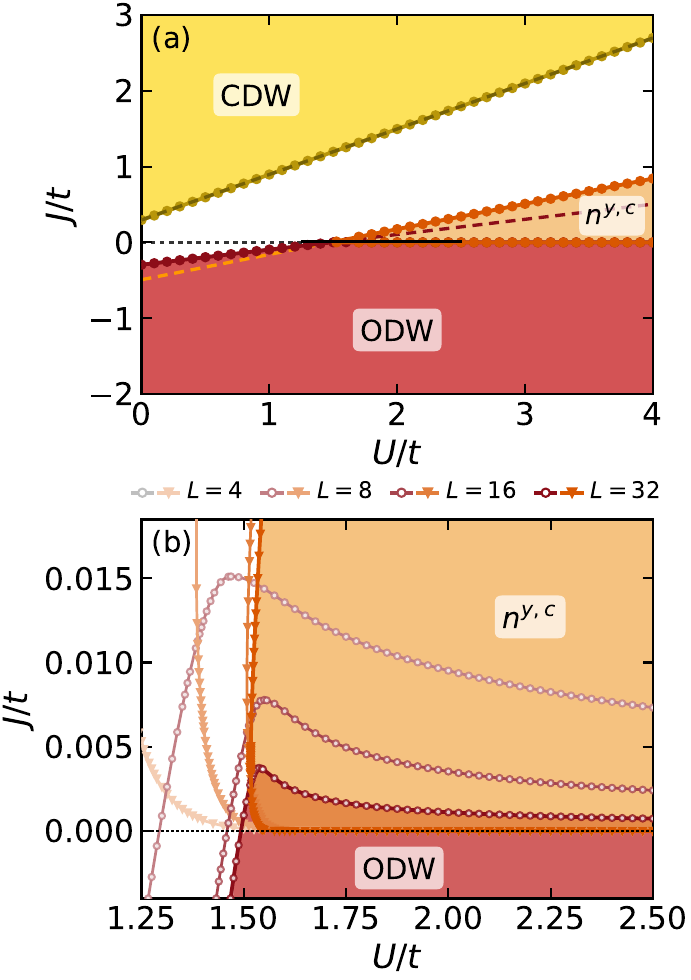}
\caption{Charge-sector competition diagram at $\beta=10/t$.
Colors indicate regions where the CDW (yellow), ODW (red), and $n^{y,c}$ (orange) phases are stable, i.e. they are the absolute minimum of the free-energy including the integration of FFs.
The ODW label represents the degenerate pair $(n^{x,c},n^{z,c})$, treated through the radial amplitude $R_c=[(\phi^{x,c})^2+(\phi^{z,c})^2]^{1/2}$.
(a) $L=32$; dashed lines mark single-channel onsets.
(b) Zoom near $J=0$, showing the same boundaries for different sizes $L=\{4,8,16,32\}$; points mark computed points.
In the overlap region (dark orange), both the ODW and $n^{y,c}$ free energies have nonzero global minima, signifying the emergence of an orbital SU(2) coexistence phase.
}
\label{fig:charge}
\end{figure}
We start our analysis of competing modes in the charge sector, namely the modes involving the unit matrix in the spin part (${\varsigma =c}$).
In Figure~\ref{fig:charge}, we map the dominant charge channels in the $U$--$J$ phase diagram: the CDW (yellow), ODW (red) and $n^{y,c}$ (orange) channels, with the corresponding single-channel onset thresholds marked as dashed lines.
We observe that the region of stability of the CDW is isolated from those of the orbital channels (ODW and $n^{y,c}$).
Therefore, we do not discuss the competition between CDW and any other phase, and we focus on the competition between orbital channels around $J=0$.
There, the ODW is favored for $J<0$ and $n^{y,c}$ for $J>0$, as suggested by the stiffness parameters (Table~\ref{tab:stiffness}), which differ by $J/2$, exchanging order at $J=0$.
We underline that, unlike the conventional CDW, these staggered phases do not involve a modulation of the total charge (density).

At $J=0$, the ODW and $n^{y,c}$ channels, corresponding to real and imaginary orbital ordering respectively, have degenerate stiffness parameters, and their matrices anticommute.
This allows rotations between the ODW and $n^{y,c}$ channels without changing the free energy (Sec.~\ref{sec:spectrum}), so fluctuation weight can be distributed between them.
Consistently with this, figure~\ref{fig:charge}(b) reveals a narrow coexistence region, in which both channels present a non-zero minima in their corresponding free energies.
This region becomes narrower with increasing lattice size $L$.
At the saddle-point level, the favored orbital channel switches from ODW to $n^{y,c}$ at $J=0$ without an energy barrier, since the fermionic contribution depends only on the total orbital amplitude (Sec.~\ref{sec:spectrum}).
The interplay between modes makes the switch between the two phases smooth with emergence of a coexistence phase, taking place over a narrow range of positive $J$, where the twofold-degenerate ODW  has a larger weight.
This coexistence window narrows with increasing $L$ and is expected to become infinitesimally narrow at $J=0^+$ in the thermodynamic limit, similarly to the SU(2) pseudospin-symmetric line of the attractive extended Hubbard model~\cite{PhysRevB.108.205156}.
Following Ref.~\cite{PhysRevB.108.205156}, we may therefore interpret the transition between ODW and $n^{y,c}$ as two distinct second-order phase boundaries.
Starting from the ODW phase, characterized by the components
$(n^{x,c},n^{z,c})$ at $J<0$, the system undergoes a second-order transition into the symmetry-driven $(n^{x,c},n^{y,c},n^{z,c})$ coexistence phase at $J=0$, where an emergent orbital SU(2) symmetry relates the three orbital components.
It is followed by a second-order transition into the $n^{y,c}$ phase for $J>0$.
\begin{figure}[!htbp]
\centering
\includegraphics[width=0.92\columnwidth]{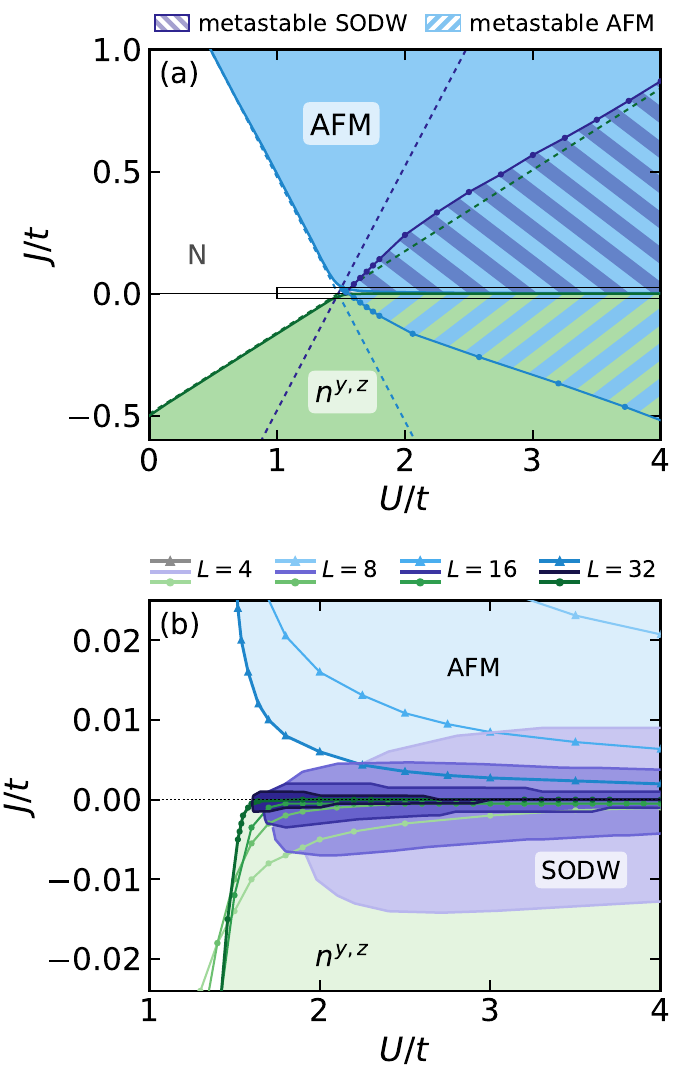}
\caption{
Collinear spin-sector phase diagram at $\beta=10/t$ (transverse spin FFs not included).
Colors indicate regions where the AFM (light blue) and $n^{yz}$ (green) phases are stable, and striped regions mark metastable configurations.
The SODW label represents the degenerate pair $(n^{x,z},n^{z,z})$ classified by $R_s$, Sec.~\ref{sec:spectrum}).
(a) $L=32$; dashed lines mark single-channel onsets.
(b) Zoom near $J=0$, showing the same boundaries for different sizes $L=\{4,8,16,32\}$ with darker colors indicate larger $L$; points mark computed points.
The SODW region (blue) lies below the AFM onset overlapping partially with the $n^{y,z}$ phase.
It narrows with increasing $L$, reaching the $J$-scan spacing $5\times10^{-4}$ at $L=32$.
}
\label{fig:spin}
\end{figure}

\begin{figure*}[!tbp]
\centering
\includegraphics[width=\textwidth]{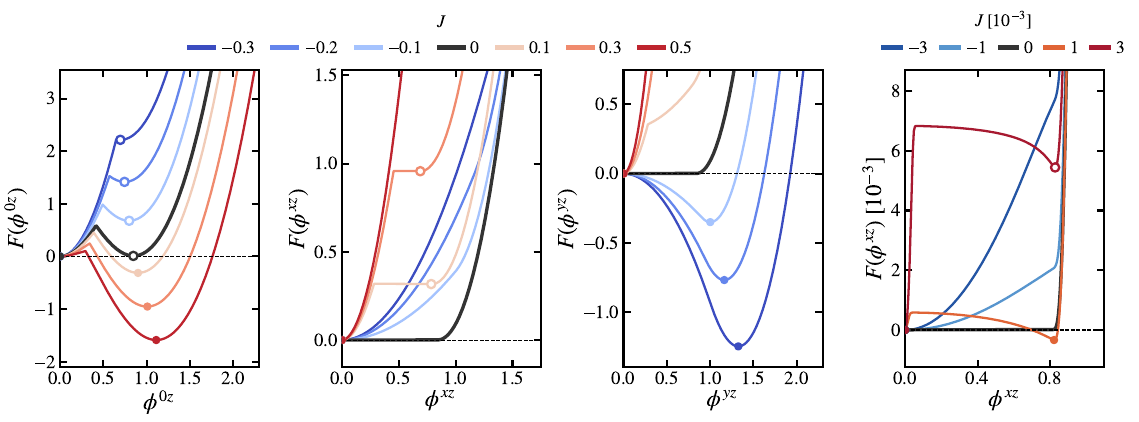}
\caption{Free energies in the magnetic sector for $U=3$, $L=32$ and $\beta=10/t$.
To better highlight the results we plot $\beta[\mathcal F_\alpha(\phi)-\mathcal F_\alpha(0)]/L^2$. 
The first three panel refer to $\phi^{0,z}$ (AFM), $\phi^{x,z}$ (SODW) and $\phi^{y,z}$ [the $\phi^{z,z}$ curve is identical to $\phi^{x,z}$].
In the fourth panel we show $\phi^{x,z}$ close to $J=0$ for $J=\{-0.003,-0.001,0,0.001,0.003\}$ in order to reveal the shallow minima and near-flat profile. 
Different values of $J$ are shown with different colors according to the legend; filled and open circles mark global and secondary minima.
Each profile integrate out the other two remaining collinear-spin FFs.
Secondary projected minimum need not be a stable point in the full spin FF space, and its height relative to zero is not an escape barrier.
}
\label{fig:spin_marginals}
\label{fig:spin_xz_smallJ}
\end{figure*}

\subsection{Competition in the spin sector}
\label{sec:res_spin}
We now consider the competition of solutions within the collinear-spin sector ($\varsigma = z$).
In Figure~\ref{fig:spin}, we map the dominant collinear-spin channels in the $U$--$J$ phase diagram: the AFM (light blue), SODW (dark blue), and $n^{y,z}$ (green), with the corresponding single-channel onset thresholds marked as dashed lines.
As discussed above, the degenerate FFs $\phi^{z,z}$ and $\phi^{x,z}$ are accounted for through their combined radial amplitude, with SODW denoting their corresponding instability.
We observe that a positive $J$ favors the orbital-homogeneous AFM, a high-spin state, while negative $J$ promotes $n^{y,z}$, a low-spin state.
This is consistent with the difference between the stiffness parameters of the two phases $J^{y,z}-J^{0,z}=J$.
As shown in Fig.~\ref{fig:spin}(b), the SODW free energy develops a nonzero global minimum only within a narrow region near $J=0$.
Whereas, the AFM and SODW phases remain separated for all system sizes studied on the positive-$J$ side, with the AFM onset consistently lying beyond the upper boundary of the SODW region, we discover a very narrow overlap region between the SODW and $n^{y,z}$ phases on the negative-$J$ side.
This signify the presence of a symmetry-driven ${(n^{x,z},n^{y,z},n^{z,z})}$ coexistence region, due to the orbital SU(2) symmetry, similar to the observation of the orbital-channel overlap in Fig.~\ref{fig:charge}(b).
As $L$ increases, the SODW and coexistence regions narrow, as the AFM and $n^{y,z}$ boundaries approaches $J=0$ from opposite sides.

The metastable regions reveal a further difference with respect to the competition in the charge sector.
A secondary minimum of the AFM free energy persists on the negative-$J$ side, while a secondary SODW minimum extends into the AFM region at positive $J$.
These regions characterized by a secondary minimum are indicated by the secondary-color stripes in Fig.~\ref{fig:spin}(a).
This behavior can be understood based on the arguments of Sec.~\ref{sec:spectrum}: the spin-orbital matrices anticommute with one another, whereas the AFM matrix commutes with both SODW matrices.
The resulting sum-and-difference gaps allow for distinct and competing free-energy minima, permitting the emergence of metastable states.
Although the AFM matrix also commutes with the matrix associated with $n^{y,z}$, at positive $J$ a configuration in this channel can lower its free energy by rotating towards the more attractive SODW sector at fixed amplitude as it crosses the $J=0$ line.
This prevents the formation of a separate metastable $n^{y,z}$ basin within the retained spin sector.

Figure~\ref{fig:spin_marginals} illustrates this behavior through the free-energy profiles of the different channels at $U=3$ and selected values of $J$.
We actually plot $\beta[\mathcal F_\alpha(\phi)-\mathcal F_\alpha(0)]/L^2$ to better highlight the difference between the various curves. 
In the region of stability of the AFM (first panel), its free energy has a global minimum at $\phi^{0,z}\neq0$.
This minimum persists as a secondary minimum over an extended range of $J\leq0$ within the region of the $n^{y,z}$ phase, indicating metastable AFM order.
A similar behavior occurs in the competition between the mutually exclusive CDW and AFM fluctuations in the single-orbital extended Hubbard model~\cite{PhysRevB.108.035143}.
A different behavior is found in the free-energy profiles of the SODW and $n^{y,z}$ channels (second and third panel).
The SODW free energy profiles develop local minima at nonzero FF over a range of $J>0$, while the $n^{y,z}$ free energy has a nonzero global minimum in the $n^{y,z}$ phase.
Near the symmetry point $J=0$, these profiles become shallow, in contrast to the AFM profile.
At $J=0$, the three spin-orbital FFs have equal stiffness parameters and their matrices anticommute, allowing rotations at fixed total amplitude without changing the free energy.
This rotational freedom explains their nearly flat free energy profiles near the symmetry point.
Such flatness does not imply a vanishing spin-orbital amplitude: for example, a three-component vector of fixed nonzero length with uniformly distributed orientations has a uniform distribution for each Cartesian component.
These free energy profiles therefore reflect fluctuations of both orientation and amplitude (SM Sec.~S4) and are related to the underlying orbital SU(2) symmetry at $J=0$.

We conclude that the transition between the imaginary spin-orbital $n^{y,z}$ phase and orbital-homogeneous AFM involves two distinct processes.
Without the AFM channel, the spin-orbital transition across $J=0$ resembles its charge-orbital counterpart: starting from SODW at $J>0$, the system undergoes a second-order transition into a symmetry-driven $(n^{x,z},n^{y,z},n^{z,z})$ phase associated with the emergent orbital SU(2) symmetry at $J=0$.
This is followed by a second-order transition into $n^{y,z}$ at $J<0$.
Including the interplay with the AFM amplitude changes the nature of the transition.
The dominant orbital-homogeneous AFM fluctuations at $J>0$ undergo a first-order transition into the regime dominated by the spin-orbital channels ${(n^{x,z},n^{y,z},n^{z,z})}$, accompanied by metastable configurations selected by $J$.
The intermediate SU(2)-symmetric phase at $J=0$ therefore connects to the AFM at $J>0$ through second- and first-order transitions in sequence, passing through the SODW phase, and to $n^{y,z}$ at $J<0$ through a second-order transition.
Metastable SODW configurations at $J>0$ and AFM configurations at $J<0$ can accompany these transitions, reflecting the mutually exclusive competition between AFM and the spin-orbital orders.

\subsection{Full competition including all channels}
\label{sec:charge_vs_spin}
\begin{figure}[b!]
\centering
\includegraphics[width=0.95\columnwidth]{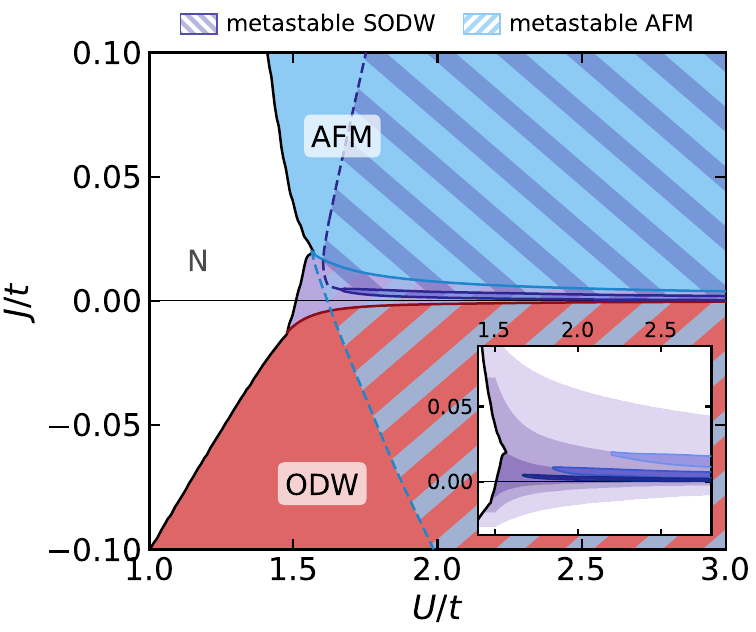}
\caption{Joint seven-FF charge--spin phase diagram at $\beta=10$ for $L=32$.
Colors indicate the stable phases according to the free-energy criterion: ODW (red), AFM (light blue) and SODW (dark blue, narrow band at small $J>0$); N denotes the normal state (white), separated from the ordered region by the solid black onset line.
In the purple region, all individual channel free energies have their global minima at zero, whereas the total-amplitude mode has a nonzero global minimum.
Striped regions mark metastable SODW and AFM (secondary marginal minima), bounded by the dashed lines.
Inset: Purple region and SODW band (dark blue outlines) for $L=\{8,16,32\}$, with darker shades for larger $L$; the black line is the $L=32$ onset.
}
\label{fig:charge_vs_spin}
\end{figure}    

As we discussed above, the saddle point hierarchy suggests that the charge and spin sectors are expected to exhibit a strong interplay near the $J=0$ and $J=U$ phase boundaries.
In particular, the region around $J=0$, where a large number of saddle-point minima are present as a result of the emergent SU(4) symmetry, calls for a theoretical treatment in which the competition between all these channels is taken into account.
As in the saddle-point analysis, the leading channels in this regime are trivially determined by the most negative stiffness parameters on their respective sides of $J=0$ (Table~\ref{tab:stiffness}).

In order to address this point we compute the MCFF phase diagram near $J=0$, retaining all seven active FFs in the region: ODW, SODW and AFM, together with the remaining orbital-hybridized $n^{x,c}$, $n^{y,c}$, $n^{x,z}$ and $n^{y,z}$ modes.
Here, the CDW mode is not included due to its stiffness parameter being positive in this region, and the spin sector is constrained to the collinear spin component ($\varsigma = s$).
We highlight that in the MCFF method, we have the possibility to study the full competition between all seven modes that become degenerate at $J=0$, and analyze their interplay, rather than restricting the fluctuations to the separate sectors.
The integral over all the different FFs is evaluated by Monte Carlo sampling~\cite{Metropolis1949,Metropolis1953}, and the free energies for each channel are reconstructed from the sampled amplitude distributions.
Each free energy for the different FF channels results from the integration over the remaining FFs, described either as doublet radii or single-FF amplitudes depending on their degeneracy.
The result is summarized in the phase diagram shown in Fig.~\ref{fig:charge_vs_spin}, where we mark with different colors the regions where individual free energies have a non-zero global-minimum.

In Fig.~\ref{fig:charge_vs_spin}, we show the $U$--$J$ phase diagram for $L=32$, identifying regions of stable and metastable order according to the free-energy analysis.
ODW (red) is favored for ${J<0}$, while AFM (light blue) is selected at sufficiently large ${J>0}$.
While these results reflect the saddle point hierarchy, in the intermediate region we find two distinctive results arising from the inclusion of the explicit interplay between the different channels.
In the purple region, we find the free energies of all the individual channels are minimized at zero amplitude, but the total-amplitude free energy has its global minimum at a finite amplitude.
Within this region we find a small sub-region where the SODW phase (blue) stabilize.
We also highlight two striped regions where metastable solutions are found in Fig.~\ref{fig:charge_vs_spin}, recalling those found in the magnetic channel: a secondary AFM minimum persisting below the stable AFM boundary, across the intervening purple regions and narrow SODW band, and into the ODW phase, while a metastable SODW state remains within the AFM-dominated region.
This similarity reflects the common structure of the fermionic spectrum.
As the ODW and SODW matrices commute with the AFM matrix, they produce branches associated with sum-and-difference gaps.
Mixing the competing amplitudes can therefore partly close a gap and separate their free-energy minima, allowing the AFM metastability to penetrate into the ODW phase.

In the inset of Fig.~\ref{fig:charge_vs_spin}, we compare the purple region and SODW boundary for $L=\{8,16,32\}$, showing that both purple branches and the SODW region narrow with increasing system size.
We can surmise that SODW phase shrinks to become infinitesimally narrow around $J=0^+$ in the thermodynamic limit, where it would be crucial to properly address the critical behavior.
We also note that the inclusions of other channels, for instance, those associated with transverse fluctuations and different momenta are expected to change our results.
Transverse fluctuations are expected to strengthen the spin-sector instabilities, yet together with different momenta, may increase the stability of the purple region where the individual channels are not stable.

\begin{figure}[t!]
    \centering
    \includegraphics[width=0.95\linewidth]{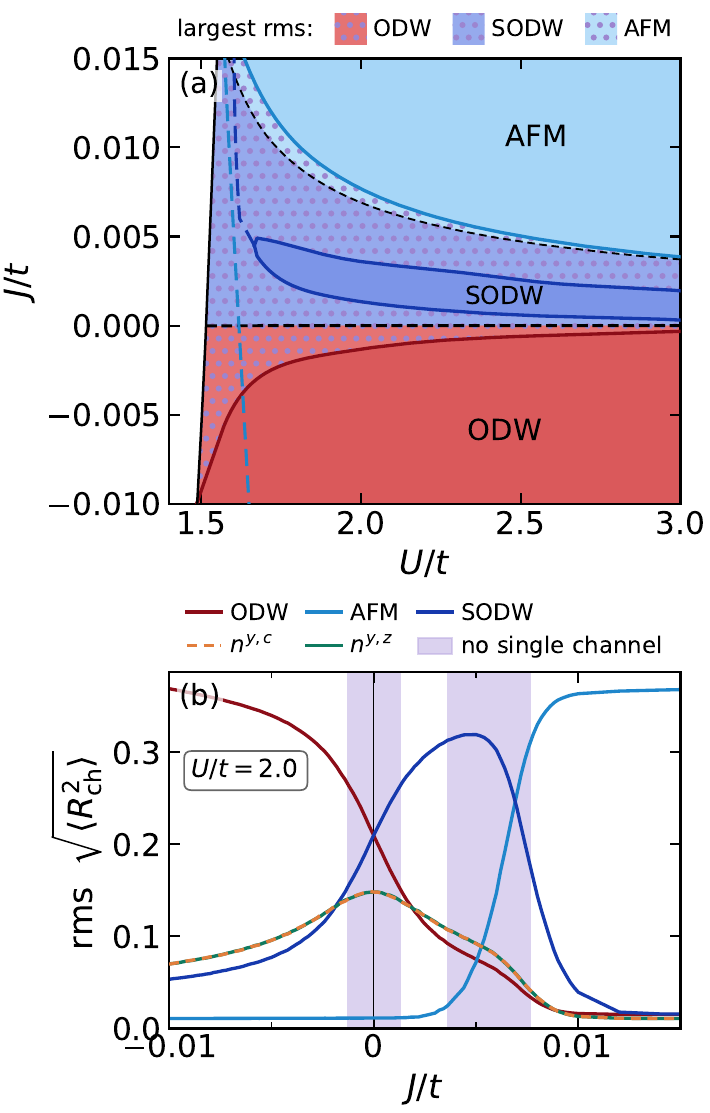}
    \caption{Dominant fluctuation channels in the joint seven-FF calculation at $L=32$ and $\beta=10$, near $J=0$.
    (a) Close-up of the phase diagram in Fig.~\ref{fig:charge_vs_spin}, focusing on the non-trivial region around $J=0$.
    Solid colors indicate phases with nonzero global free-energy minima, while dotted regions correspond to the purple regions of Fig.~\ref{fig:charge_vs_spin}, where all individual channel free energies are minimized at zero.
    The colors of the dotted areas identify the channel with the largest rms amplitude $\sqrt{\langle R_{\rm ch}^2\rangle}$.
    Dashed black lines mark equal rms amplitudes between the orbital-charge and spin-orbital sectors at $J=0$, and between the AFM and spin-orbital doublet at positive $J$.
    Colored dashed lines mark the metastability boundaries shown in Fig.~\ref{fig:charge_vs_spin}.
    The solid black line marks the numerical onset boundary; white denotes the normal-state region.
    (b) rms amplitudes versus $J$ at $U=2$; shaded bands correspond to the dotted regions in (a).
    The $n^{y,c}$ and $n^{y,z}$ curves overlap.}
\label{fig:dominant_channel}
\end{figure}

In order to gain further insight and to identify which channels carry the strongest fluctuations across the phase diagram, in Fig.~\ref{fig:dominant_channel}, we superimpose onto the phase diagram near $J=0$, the root-mean-square (rms) amplitudes of the competing channels at $L=32$.
These are obtained from the Monte Carlo sampling of the joint seven-FF distribution as $\sqrt{\langle R_{\rm ch}^2\rangle}$, where $R_{\rm ch}^2$ is the sum of the squared amplitudes of the FFs belonging to each channel. 
While the channel with the largest rms amplitude contributes most strongly to the mean-square FF amplitude, this does not imply that it dominates individual configurations.

In panel (a) of Fig. ~\ref{fig:dominant_channel}, we plot this information in dotted regions, where the color of the dotted areas signify the channel with the largest rms amplitude.
Here, we use the same color legend as used to draw the phase diagram.
On this scale, the dotted regions between the AFM and ODW phases, i.e. the purple regions in Fig.~\ref{fig:charge_vs_spin}, and the narrow SODW band separating them, are clearly resolved.
The solid colored boundaries follow the free-energy-minimum criterion of Fig.~\ref{fig:charge_vs_spin}, while the dashed black lines mark where the rms amplitudes of the corresponding channels become equal.
The latter therefore indicate crossovers in fluctuation weight rather than any additional ordering transitions.
Consistent with the prior analysis, the ODW fluctuations have the largest rms amplitude for ${J<0}$, while the SODW fluctuations take over for small positive $J$.
The crossover in the fluctuations seems to take place exactly at the symmetry line $J=0$. 
Increasing $J$ further, AFM fluctuations overtake the SODW fluctuations with a crossover, which appears to be located slightly before the AFM phase boundary as defined by the free-energy-minimum criterion.
This comparison accounts for the combined amplitude of the two components of each doublet, whereas AFM is represented by a single FF.
Interestingly, the $\{n^{y,c}, n^{y,z}\}$ components never carry the largest rms amplitude in the joint calculation, despite being favored in their respective portions of the phase diagrams obtained in the charge and spin sectors (Secs.~\ref{sec:res_charge} and \ref{sec:res_spin}).

Figure~\ref{fig:dominant_channel}(b) illustrates the redistribution as a function of $J$ along a cut at $U=2$, with the shaded bands marking the two dotted regions crossed in panel (a).
At $J=0$, the six retained components involving orbital fluctuations have comparable rms amplitudes, reflecting the underlying SU(4) symmetry, whereas the AFM fluctuations (blue) are much weaker.
This imbalance is obtained within the restricted seven-FF ensemble, whose limitations are discussed later.
Moving away from $J=0$, the fluctuation weight shifts towards the ODW or SODW regions according to the sign of $J$.
On the positive-$J$ side, the growth of the SODW amplitude is limited by the sharp rise of the AFM fluctuations as the AFM phase is approached.
As a general result, the competing amplitudes do not change sharply  across the free-energy-minimum boundaries, remaining sizeable.
It shows that stabilization of one channel does not eliminate fluctuations in the others.
We highlight, that at $J=0$, the six degenerate components have approximately equal mean-square amplitudes, as the rms amplitude for each doublet is larger by a factor $\sqrt{2}$ relative the individual components.
Additionally, we observe the subleading ${(n^{y,c}, n^{y,z})}$ components are degenerate with a maximal rms value at $J=0$.
Note that rms amplitudes cannot alone distinguish competition between configurations dominated by different channels from configurations in which several channels simultaneously have appreciable amplitudes.

We underline that the results here depend on the choice of retained FFs included in our calculations, and that neglecting some FFs may limit the ability of the method to incorporate interaction effects beyond the saddle-point starting point.
For example, the seven-FF ensemble does not display the full SU(4) symmetry, with the expected SU(4) coexistence phase at $J=0$ being substituted by a six-component symmetry-driven ${(n^{x,c/z},n^{y,c/z},n^{z,c/z})}$ coexistence phase associated with the orbital SU(2) symmetry.
Its emergence relies on the existence of "pathways", where all pairs of generator matrices at least share a common generator matrix which they mutual commute with.
These pathways permit the redistribution of fluctuation weight at zero free-energy cost at $J=0$.
As the AFM generator $\tau^0\otimes\sigma^z$ commutes with the six retained generators involving orbital fluctuations in the joint seven-FF calculation, coexistence between their modes is forbidden.
Including transverse components, however, despite not affecting the saddle-point phase diagram, these FFs are expected to reshape the interacting phase diagram by permitting new pathways for commutation between FF modes, absent in the collinear ensemble.
By retaining all FF spin sector modes, we expect the AFM--SODW phase boundary to become characterized rather by a second-order transition, as, e.g., the collinear AFM mode $n^{c,z}$ anticommutes with the transverse SODW modes ${(n^{x,x},n^{y,x},n^{x,y},n^{y,y})}$.
Moreover, we expect their inclusion to strengthen the FFs of the spin-sector, and the emergence of the SU(4) coexistence phase, related to all fifteen traceless particle--hole channels modes being degenerate at $J=0$.
This is expected to favor the "purple" region near $J=0$, where the individual channels have zero amplitude.

Finally, we briefly discuss the more transparent CDW--AFM phase boundary at $J=U$, that involves only two components in the particle-hole channel. As their corresponding matrices commute, they exhibit a first-order transition following the $J=U$ line, joined by secondary metastable configurations.
This boundary behaves nearly identically to the CDW--AFM phase boundary found in the half-filled single-orbital extended Hubbard model~\cite{PhysRevB.108.035143}.

\section{Conclusions}
\label{sec:conclusions}
We have developed a multiorbital extension of the MCFF theory that provides a computationally inexpensive framework for exploring competing instabilities of strongly correlated electron models and their phase diagrams.
Building on the single-orbital formulation~\cite{PhysRevB.108.035143,PhysRevB.108.205156}, our general approach treats charge, spin and orbital channels within a common variational construction, and can easily be extended to include superconducting channels.
Integrating out the fermions reduces the problem to an integral over selected static collective FFs, retaining their coupled, non-Gaussian thermal fluctuations beyond a mean-field saddle point.
Direct access to their free-energy profiles allows competing global and local minima to be identified within the retained ensemble, making the method well suited to systematic parameter scans and finite-size comparisons.

We tested the method on the half-filled two-dimensional two-orbital Kanamori--Hubbard model~\cite{Kanamori1963,Werner2007,PhysRevB.79.245128,PhysRevB.81.035112,PhysRevB.83.125110,Werner2orb,Capone2orb,PhysRevLett.129.096404,PhysRevLett.132.226501,Florez,Torchia,Soares2026}, where the Hubbard interaction and the Hund's coupling shape the competition between several phases. 
The accuracy of the method relies on the number and nature of the chosen modes to be included. The MCFF construction may in principle include all the possible amplitudes and fluctuations. 
In this work, we restricted ourselves to the particle-hole channels with commensurate momentum $\Qv=\Piv$.
This choice is specific to the present application, motivated by the perfect nesting of the bare dispersion $\epsilon_{\kv}$.

With our choice to include all the charge and spin-collinear phases with $\Qv=\Piv$, we find the region in the vicinity of $J=0$, associated with a SU(4) symmetry in our model, to exhibit an intricate competition between collective charge, spin and orbital fluctuations.
Interestingly, we discover the inclusion of the various FFs leads to a region where the collective FF is non-zero, despite none of the individual FFs displays a finite amplitude due to the strong interplay.
This region marks a clear qualitative difference with respect to the saddle-point solution, demonstrating the ability of the method to depart from mean-field theories.
We stress that this region of strong fluctuations shrinks with the system size, and while we can not draw definitive conclusions about the thermodynamic limit, we expect their importance in describing the critical effects occurring at the crossing of $J=0$ to survive.
Moreover, one can expect that the inclusion of some of the neglected modes can modify as well as expand this region.
A systematic study of these effects including in a controlled way is a promising avenue for future investigations.
Another intriguing direction could be to combine the FF theory with non-conventional mean-field solutions, such as stripes~\cite{Schulz,Metzner} or more involved spin patterns~\cite{Quinzi2026}.

This non-trivial example demonstrates the potential of the MCFF method, allowing for a controlled and efficient description for investigating competing collective orders.
Nevertheless, dynamical correlations and omitted channels and momenta can affect quantitative phase boundaries.
These questions are expected to be particularly relevant in regions of strong competition and in the strongly correlated regime. 
Yet, we believe the flexibility of the MCFF method and ability to disentangle various contributions constitute a potent diagnostic tool.
Thus, applying the framework to models with more than two orbitals, including additional fluctuating channels and considering more general band structures would broaden its range of applications.

\begin{acknowledgments}
Authors acknowledge financial support from MUR through the PRIN 2020 (Prot. G93C22000430006), from National Recovery and Resilience Plan (PNRR) MUR (Prot. PE0000023-NQSTI), financed through Next Generation EU by the European Union, and from FVG through the Complementary Operational Program (POC) 2014-2020 (Prot. G93C25000880002).
High-performance calculations were carried out thanks to resources provided by CINECA HPC via Convenzione SISSA 2025 and 2026.
\end{acknowledgments}

\bibliography{bib_file}

\onecolumngrid
\clearpage
\begingroup
\setcounter{section}{0}\setcounter{subsection}{0}
\setcounter{equation}{0}\setcounter{table}{0}\setcounter{figure}{0}
\renewcommand{\thesection}{S\arabic{section}}
\renewcommand{\theequation}{S\arabic{equation}}
\renewcommand{\thetable}{S\arabic{table}}
\renewcommand{\thefigure}{S\arabic{figure}}
\renewcommand{\theHsection}{S\arabic{section}}
\renewcommand{\theHequation}{S\arabic{equation}}
\renewcommand{\theHtable}{S\arabic{table}}
\renewcommand{\theHfigure}{S\arabic{figure}}
\renewcommand{\kv}{\mathbf{k}}
\renewcommand{\Qv}{\mathbf{Q}}
\renewcommand{\Piv}{\mathbf{M}}
\renewcommand{\id}{\mathbb I}
\renewcommand{\Se}{\mathcal S_e}
\providecommand{\Sphi}{}\renewcommand{\Sphi}{\mathcal S_\phi}
\renewcommand{\eve}[1]{\langle #1\rangle_{\Se}}

\begin{center}
{\large\bfseries Supplemental Material\\[0.3cm]
Investigating competing order in the Kanamori-Hubbard model\\
with the multiorbital multichannel fluctuating field theory\par}
\vspace{0.4cm}
L. Torchia$^{1}$, E. Linn\'er$^{1}$, M. Capone$^{1,2}$\\[0.2cm]
{\footnotesize $^{1}$International School for Advanced Studies (SISSA), Via Bonomea 265, 34136 Trieste, Italy}\\
{\footnotesize $^{2}$CNR-IOM, Istituto Officina dei Materiali, Consiglio Nazionale delle Ricerche, Via Bonomea 265, 34136 Trieste, Italy}
\end{center}
\vspace{0.4cm}

In these supplemental material, we present the detailed derivation for the multiorbital multichannel fluctuating field (MCFF) construction for charge and spin fluctuations in the particle-hole channel of the two-orbital Kanamori--Hubbard model, employed in the main text.
First, we determine the channel stiffness parameters from a variational principle, by evaluating the local interaction contractions.
Second, we present the free energy construction, giving detailed account for the single-channel onset and the different fluctuating-field (FF) channel free-energy profiles describing the interplay between collective instabilities.
Finally, we discuss the symmetry relations underlying the competition near the phase boundaries $J=0$ and $J=U$.

\section{Variational construction}
\label{SM:sec:variational}

Our multiorbital MCFF construction is built on the same variational principle, as utilized in the single-orbital FF and MCFF constructions~\cite{SM:PhysRevB.102.224423,SM:PhysRevB.108.035143,SM:PhysRevB.108.205156}.
For this purpose, consider the original lattice action $\mathcal S=\mathcal S_0+\mathcal S_{\mathrm{int}}$, introduced in the main text, with
\begin{equation}
 \mathcal S_0=-\frac{1}{\beta N_s}\sum_{k,m,\sigma}
 c^*_{km\sigma}\mathcal G_k^{-1}c_{km\sigma},
 \hspace*{0.3cm}\text{and}\hspace*{0.3cm}
 \mathcal S_{\mathrm{int}}=\sum_j\int_0^\beta d\tau\,H_{\mathrm{int}}(j,\tau).
 \label{SM:eq:action}
\end{equation}
Here, $k=(\kv,\nu)$ labels the combined quasi-momentum $\kv$ and fermionic Matsubara frequency $\nu$, and $N_s$ is the number of sites, whereas $H_{\mathrm{int}}$ is the local Kanamori interaction, with all defined in the main text.
The aim of the MCFF theory is to map the original lattice action onto the MCFF action, which couples the various densities to the a set of retained FFs:
\begin{equation}
 \mathcal S^*=\mathcal S_0+
 \sum_{\alpha,\Qv}\left[
 \phi^\alpha_{\Qv}\rho^\alpha_{-\Qv}
 -\frac{\beta N_s}{2J^\alpha_{\Qv}}
 \phi^\alpha_{\Qv}\phi^\alpha_{-\Qv}\right].
 \label{SM:eq:trial}
\end{equation}
We consider a set of retained FFs which coupled to the particle--hole channels, as defined by $\Gamma^\alpha=\tau^\varrho\otimes\sigma^\varsigma$, where $\alpha=(\varrho,\varsigma)$ denotes the orbital $\varrho\in\{0,x,y,z\}$ and spin $\varsigma\in\{c,x,y,z\}$ textures.
In addition, we use the shifted densities $\rho^\alpha_{\Qv}$, ensuring the chemical potential $\mu$ is identical between the original lattice action and MCFF action.
As the real-field integral involving $\mathcal S^*$ only converges for stiffness parameters satisfying $J^\alpha_{\Qv}<0$, defining a stability criterion for the retained collective instabilities.

Integrating out the FFs explicitly from the MCFF action, gives an effective fermionic action,
\begin{equation}
 \mathcal S_c=\mathcal S_0+
 \frac12\sum_{\alpha,\Qv}\frac{J^\alpha_{\Qv}}{\beta N_s}
 \rho^\alpha_{\Qv}\rho^\alpha_{-\Qv},
 \label{SM:eq:Sc}
\end{equation}
which forms the basis for determining the stiffness parameters with the Peierls--Feynman--Bogoliubov variational principle~\cite{SM:PhysRev.54.918,SM:Bogolyubov:1958zv,SM:feynman1972}.
The variational principle is based on the functional:
\begin{align}
 f[J^\alpha_{\Qv}] & \equiv f_{c}[J^\alpha_{\Qv}]
 +\frac{1}{\beta N_s}\left\langle
 \mathcal S- \mathcal S^* \right\rangle_{\mathcal S_c} = -\frac{\ln\mathcal Z_c}{\beta N_s}
 +\frac{1}{\beta N_s}\left\langle
 \mathcal S_{\mathrm{int}}-
 \frac12\sum_{\alpha,\Qv}\frac{J^\alpha_{\Qv}}{\beta N_s}
 \rho^\alpha_{\Qv}\rho^\alpha_{-\Qv}
 \right\rangle_{\mathcal S_c},
 \label{SM:eq:Fvar}
\end{align}
given per site.
In evaluating the interaction average in Eq.~\eqref{SM:eq:Fvar}, we exploit the following exact relationship:
\begin{equation}
 \langle O\rangle_{\mathcal S_c}
 =\big\langle\eve{O}\big\rangle_{\Sphi},
 \hspace*{0.3cm}\text{where}\hspace*{0.3cm}
 \Se=\mathcal S_0+\sum_{\alpha,\Qv}\phi^\alpha_{\Qv}\rho^\alpha_{-\Qv}.
 \label{SM:eq:nested}
\end{equation}
Firstly, the inner average describes noninteracting fermions for fixed FFs, so its local four-fermion contractions follow conveniently and exactly from Wick's theorem, as performed in the next section.
Secondly, the outer average accounts for the distribution of the FFs, defined by the FF distribution obtained from the effective FF action, where the fermionic degrees of freedom have been explicitly integrating out.
For the collective-density product in Eq.~\eqref{SM:eq:Fvar}, we follow the single-orbital construction and retain only the disconnected contribution at fixed fields:
\begin{equation}
 \big\langle\rho^\alpha_{\Qv}\rho^\alpha_{-\Qv}\big\rangle_{\mathcal S_c}
 \simeq
 \left\langle\eve{\rho^\alpha_{\Qv}}
 \eve{\rho^\alpha_{-\Qv}}\right\rangle_{\Sphi}.
 \label{SM:eq:factorization}
\end{equation}
It is motivated by the omitted connected contribution being subleading in the system size.
This approximation concerns only the collective-density product, with the local Wick contractions themselves remaining exact at fixed FFs.

After projection onto the retained modes, both interaction contributions in Eq.~\eqref{SM:eq:Fvar} have quadratic forms in the FF channel amplitudes.
It permits from the variational prescription to give the stiffness parameters $J^\alpha=2C_\alpha$, with the local interaction average having the coefficient $C_\alpha$ multiplying $\eve{n^\alpha}^2$.
Thus, the stiffness parameters follow from the interaction contractions without any additional choice of Fierz decoupling.
We highlight, that stiffness parameter are independent of the quasi-momentum due to the interaction being local.
In the following section, we will analytically evaluate the form of the different stiffness parameters as a function of the intra-orbital and inter-orbital interactions parameters $U$ and $U'$, respectively, and Hund's coupling $J$.

\section{Kanamori contractions and stiffness parameters}
\label{SM:sec:stiffness}

For the variational determination of the stiffness parameters, we require to evaluate the local interaction contractions, where we exploit an explicit usage of Wick's theorem, as discussed in the previous section.
For convenience, we keep the five interaction terms separate:
\begin{equation}
 \hat H_{\mathrm{int}}=U\hat H_{1}+U'\hat H_{2}+(U'-J)\hat H_{3} + (-J)\hat H_{4}+J\hat H_{5}.
 \label{SM:eq:Hsplit}
\end{equation}
Suppressing their site index, the normal-ordered interaction terms take the form:
\begin{align}
\hat H_{1} &=\sum_m\hat c^\dagger_{m\uparrow}\hat c^\dagger_{m\downarrow} \hat c_{m\downarrow}\hat c_{m\uparrow},\nonumber\\
\hat H_{2} &=\sum_{m\ne m'}\hat c^\dagger_{m\uparrow}\hat c^\dagger_{m'\downarrow} \hat c_{m'\downarrow}\hat c_{m\uparrow},\nonumber\\
\hat H_{3} &=\sum_\sigma\hat c^\dagger_{1\sigma}\hat c^\dagger_{2\sigma} \hat c_{2\sigma}\hat c_{1\sigma},\\
\hat H_{4} &=-\sum_{m\ne m'}\hat c^\dagger_{m\uparrow}\hat c^\dagger_{m'\downarrow} \hat c_{m\downarrow}\hat c_{m'\uparrow},\nonumber\\
\hat H_{5} &=\sum_{m\ne m'}\hat c^\dagger_{m\uparrow}\hat c^\dagger_{m\downarrow} \hat c_{m'\downarrow}\hat c_{m'\uparrow}.\nonumber
 \label{SM:eq:Hterms}
\end{align}
Here the minus sign in the normal-ordered spin-flip term $\hat H_{4}$ originates from exchanging a creation and an annihilation operator, with its prefactor in Eq.~\eqref{SM:eq:Hsplit} remaining as $-J$.
Likewise, the pair hopping term $\hat H_{5}$ retains its prefactor $+J$.

Limiting our considerations to the particle-hole channel, anomalous contractions vanish and Wick's theorem gives us the following exact identity:
\begin{equation}
 \eve{\hat c^\dagger_\mu\hat c^\dagger_\nu\hat c_\rho\hat c_\lambda}
 =G_{\mu\lambda}G_{\nu\rho}-G_{\mu\rho}G_{\nu\lambda}.
 \label{SM:eq:wick}
\end{equation}
Here, we introduce $G_{\mu\nu}=\eve{\hat c^\dagger_\mu\hat c_\nu}$ for fixed FFs, with $\{ \mu,\nu \}$ denoting combined $(m,\sigma)$ components.
In the following, for notational convenience, $\langle\ldots\rangle$ will denote the expectation value with respect to $\langle\cdot\rangle_{\Se}$ at fixed FFs.
Moreover, it useful in our evaluation of the local interaction contractions to highlight that the orbital-resolved averages ${\langle n^b_{mm'}\rangle=\sum_{\sigma\sigma'}\sigma^b_{\sigma\sigma'}G_{m\sigma,m'\sigma'}}$ are related to the charge and spin channels:
\begin{equation}
 \begin{aligned}
 \langle n^b_{11}\rangle&=\tfrac12\big(\langle n^{0,b}\rangle+\langle n^{z,b}\rangle\big),&
 \langle n^b_{22}\rangle&=\tfrac12\big(\langle n^{0,b}\rangle-\langle n^{z,b}\rangle\big),\\
 \langle n^b_{12}\rangle&=\tfrac12\big(\langle n^{x,b}\rangle+i\langle n^{y,b}\rangle\big),&
 \langle n^b_{21}\rangle&=\tfrac12\big(\langle n^{x,b}\rangle-i\langle n^{y,b}\rangle\big),
 \end{aligned}
 \label{SM:eq:orbital}
\end{equation}
employing unshifted densities.

By applying the identities introduced, we may evaluate the averages of the interaction terms.
For the intra-orbital interaction, the Hartree and Fock contractions give
\begin{align}
 \eve{\hat H_{1}}
 &=\sum_m\left(G_{m\uparrow,m\uparrow}G_{m\downarrow,m\downarrow}
 -G_{m\uparrow,m\downarrow}G_{m\downarrow,m\uparrow}\right)
 =\frac14\sum_m\left[\langle n^c_{mm}\rangle^2-\sum_{s=x,y,z}\langle n^s_{mm}\rangle^2\right]\nonumber\\
 &=\frac18\left[\langle n^{0,c}\rangle^2+\langle n^{z,c}\rangle^2
 -\sum_{s=x,y,z}(\langle n^{0,s}\rangle^2+\langle n^{z,s}\rangle^2)\right].
 \label{SM:eq:HU}
\end{align}
Applying the same substitutions to the two inter-orbital density terms yields:
\begin{align}
 \eve{\hat H_{2}}&=\frac18\Big[
 \langle n^{0,c}\rangle^2-\langle n^{z,c}\rangle^2-\langle n^{0,z}\rangle^2+\langle n^{z,z}\rangle^2
 -\sum_{s=x,y}\big(\langle n^{x,s}\rangle^2+\langle n^{y,s}\rangle^2\big)\Big],\label{SM:eq:HUprime}\\
 \eve{\hat H_{3}}&=\frac18\Big[
 \langle n^{0,c}\rangle^2-\langle n^{z,c}\rangle^2+\langle n^{0,z}\rangle^2-\langle n^{z,z}\rangle^2 -\langle n^{x,c}\rangle^2-\langle n^{y,c}\rangle^2-\langle n^{x,z}\rangle^2-\langle n^{y,z}\rangle^2\Big].
 \label{SM:eq:Hparallel}
\end{align}
The remaining spin-flip and pair hopping terms take the form:
\begin{align}
 \eve{\hat H_{4}}&=\frac18\Big[
 \sum_{s=x,y}\big(\langle n^{0,s}\rangle^2-\langle n^{z,s}\rangle^2\big)
 -\langle n^{x,c}\rangle^2-\langle n^{y,c}\rangle^2+\langle n^{x,z}\rangle^2+\langle n^{y,z}\rangle^2\Big],\label{SM:eq:Hsf}\\
 \eve{\hat H_{5}}&=\frac18\Big[
 \langle n^{x,c}\rangle^2-\langle n^{y,c}\rangle^2-\sum_{s=x,y,z}\big(\langle n^{x,s}\rangle^2-\langle n^{y,s}\rangle^2\big)\Big].
 \label{SM:eq:Hph}
\end{align}
Thus, we observe pair hopping contributions to the charge and spin channels, despite no pairing channel being retained.
To collect the contributions of each coupling, we separate the prefactor $U'-J$ of $\hat H_{3}$ into its $U'$ and $J$ parts, and combine the latter with the spin-flip and pair-hopping contributions.
This gives, from Eqs.~\eqref{SM:eq:HU}--\eqref{SM:eq:Hph}:
\begin{align}
 U:\phantom{*} &\frac{U}{8}\Big[
 \langle n^{0,c}\rangle^2+\langle n^{z,c}\rangle^2
 -\sum_{s=x,y,z}\big(\langle n^{0,s}\rangle^2+\langle n^{z,s}\rangle^2\big)\Big], \nonumber
 \\[1ex]
 U':\phantom{*} &\frac{U'}{8}\Big[
 2\langle n^{0,c}\rangle^2-2\langle n^{z,c}\rangle^2
 -\langle n^{x,c}\rangle^2-\langle n^{y,c}\rangle^2-\sum_{s=x,y,z}\big(\langle n^{x,s}\rangle^2+
 \langle n^{y,s}\rangle^2\big)\Big],
 \\[1ex]
 J:\phantom{*} &\frac{J}{8}\Big[
 -\langle n^{0,c}\rangle^2+\langle n^{z,c}\rangle^2
 +3\langle n^{x,c}\rangle^2+\langle n^{y,c}\rangle^2 +\sum_{s=x,y,z}\big(-\langle n^{0,s}\rangle^2
 -\langle n^{x,s}\rangle^2+\langle n^{y,s}\rangle^2
 +\langle n^{z,s}\rangle^2\big)\Big]. \nonumber
\label{SM:eq:coupling_contributions}
\end{align}
Their sum gives $\eve{\hat H_{\mathrm{int}}}$.
In this form, all three spin components enter with the same coefficient, making the spin isotropy in the full interaction, despite a selection of separate Hamiltonian terms breaking this symmetry.
Matching twice the coefficient of each squared channel average to its stiffness parameter, as states in Sec.~\ref{SM:sec:variational}, gives us:
\begin{equation}
 \eve{\hat H_{\mathrm{int}}}
 =\frac12\sum_{\varrho=\{0,x,y,z\}}\left[
 J^{\varrho,c}\langle n^{\varrho,c}\rangle^2+J^{\varrho,s}\sum_{s=\{x,y,z\}}\langle n^{\varrho,s}\rangle^2\right],
 \label{SM:eq:diagonal}
\end{equation}
where the $J^{\varrho,s}$ parameters are independent of the spin direction $s\in\{x,y,z\}$.
For example, we obtain the orbital-homogeneous charge and spin channels are given by $J^{0,c}=(U+2U'-J)/4$ and $J^{0,s}=-(U+J)/4$, respectively.
All resulting stiffness parameters are listed in Table~\ref{SM:tab:stiffness}.
It is important to highlight that the diagonal form does not decouple the instability channels, which remain coupled through the fermionic determinant.

\begin{table}[htbp]
\centering
\caption{Particle--hole stiffness parameters from the local Kanamori contractions.
Spin label $s \in \{x,y,z \}$ denote its direction, with the calculations in the main text retaining only $s=z$.
Rotationally invariant relation $U'\equiv U-2J$ is employed in the last column.
}
\label{SM:tab:stiffness}
\begin{tabular}{@{}llll@{}}
\toprule
Sector & Channel & $J^{a,b}$ & $J^{a,b}$ for $U'=U-2J$\\
\midrule
Charge & $n^{0,c}$ & $(U+2U'-J)/4$ & $(3U-5J)/4$\\
 & $n^{x,c}$ & $(-U'+3J)/4$ & $(-U+5J)/4$\\
 & $n^{y,c}$ & $(-U'+J)/4$ & $(-U+3J)/4$\\
 & $n^{z,c}$ & $(U-2U'+J)/4$ & $(-U+5J)/4$\\
\midrule
Spin & $n^{0,s}$ & $-(U+J)/4$ & $-(U+J)/4$\\
 & $n^{x,s}$ & $-(U'+J)/4$ & $(-U+J)/4$\\
 & $n^{y,s}$ & $(-U'+J)/4$ & $(-U+3J)/4$\\
 & $n^{z,s}$ & $(-U+J)/4$ & $(-U+J)/4$\\
\bottomrule
\end{tabular}
\end{table}

Focusing on the rotationally invariant relation $U'\equiv U-2J$ employed in the main text, the equalities $J^{x,c}=J^{z,c}$ and $J^{x,s}=J^{z,s}$ reflect the orbital rotation generated by $\tau^y$.
Moreover, at $J=0$, all fifteen traceless channels have stiffness $-U/4$, whereas $J^{0,c}=3U/4$.
This is the channel representation of the SU(4)-symmetric interaction $\hat H_{\mathrm{int}}=U\hat N(\hat N-1)/2$ occurring at $J=0$.

\section{Free energy and single-channel onset}
\label{SM:sec:freeenergy}

In the MCFF theory, a free energy construction can be developed on the basis of the effective action for the FFs degrees of freedom, given in the main text as:
\begin{equation}
    \mathcal{S}_{\phi} = - \mathrm{Tr}\ln\Big[ \mathcal{G}^{-1} - \sum_{\alpha,\Qv} \phi^{\alpha}_{\Qv} \Lambda^{\alpha}_{-\Qv} \Big] - \frac{1}{2}\sum_{\alpha,\Qv} \frac{\beta N_s}{J^{\alpha}_{\Qv}}\, \phi^{\alpha}_{\Qv} \phi^{\alpha}_{-\Qv}.
\end{equation}
Underlying the fermionic degrees of freedom in the free energy are thus described by an internal noninteracting Hamiltonian $H_\kv$ in the $- \mathrm{Tr}\ln\big[ \dots \big]$ term.
Consider at half filling the components of the degenerate bands described by the bare dispersion $\epsilon_\kv$, as in the main text, coupled by a fixed generic FF configuration with ordering $\Qv=(\pi,\pi)$, the Hamiltonian $H_\kv$ takes them the following form describing the internal behavior:
\begin{equation}
 H_\kv=\begin{pmatrix}
 \epsilon_\kv\id_4&\Phi\\
 \Phi&-\epsilon_\kv\id_4
 \end{pmatrix},
 \hspace*{0.3cm}\text{with}\hspace*{0.3cm}\Phi=\sum_\alpha\phi^\alpha\Gamma^\alpha.
 \label{SM:eq:Hk}
\end{equation}
If $\lambda_i$, $i=\{1,\ldots,4\}$, are the eigenvalues of $\Phi$, the fermionic energies are described by a spectrum given by $\pm E_{\kv i}$, with $E_{\kv i}=\sqrt{\epsilon_\kv^2+\lambda_i^2}$.
This leads to the free energy. up to a FF-independent constant:
\begin{equation}
 \beta\mathcal F(\{\phi\})=
 -\sum_{\kv\in\mathrm{RBZ}}\sum_{i=1}^4
\ln\!\left[2+2\cosh\!\left(\beta\sqrt{\epsilon_\kv^2+\lambda_i^2}\right)\right]
 -\frac{\beta N_s}{2}\sum_\alpha\frac{(\phi^\alpha)^2}{J^\alpha},
 \label{SM:eq:freeenergy}
\end{equation}
which is express for a single FF, without any interplay with other FFs.
Here, the reduced Brillouin zone (RBZ) contains one momentum from each pair $(\kv,\kv+\Qv)$, and each logarithm already includes the two levels $\pm E_{\kv i}$.
For the retained negative stiffness parameters, the final term is always positive.

The spectrum related to the internal Hamiltonian can be expressed directly in terms of the FFs used in our calculations.
Defining the orbital vectors
\begin{equation}
 \boldsymbol\phi_c=(\phi^{x,c},\phi^{y,c},\phi^{z,c}),
 \hspace*{0.3cm}\text{and}\hspace*{0.3cm}
 \boldsymbol\phi_z=(\phi^{x,z},\phi^{y,z},\phi^{z,z}),
 \label{SM:eq:vectors}
\end{equation}
we obtain in the charge sector that the FF eigenvalues are $\phi^{0,c}\pm|\boldsymbol\phi_c|$, each twice degenerate, and in the collinear spin sector they are $\pm\phi^{0,z}\pm|\boldsymbol\phi_z|$.
For the joint seven-field ensemble, as defined in the main text with the $\phi^{0,c}$ absent, the spin remains block diagonal and the four eigenvalues are given by:
\begin{equation}
 \lambda_i\in\left\{
 \phi^{0,z}\pm|\boldsymbol\phi_c+\boldsymbol\phi_z|,
 - \phi^{0,z}\pm|\boldsymbol\phi_c-\boldsymbol\phi_z|
 \right\}.
 \label{SM:eq:seven}
\end{equation}
Therefore, the relative orientation of the two orbital vectors only enters through the joint free energy.

For a single FF the situation is significantly simpler.
As $(\Gamma^\alpha)^2=\id_4$ implies $\lambda_i^2=\phi^2$ in every channel, we may treat all single FF modes identically.
Expanding Eq.~\eqref{SM:eq:freeenergy} around zero gives
\begin{equation}
 \beta[\mathcal F(\phi)-\mathcal F(0)]
 =\beta N_s\left[\frac{1}{2|J^\alpha|}-2\chi_0(L)\right]\phi^2
 +O(\phi^4),
 \hspace*{0.3cm}\text{with}\hspace*{0.3cm}
 \chi_0(L)=\frac{1}{N_s}\sum_{\kv\in\mathrm{BZ}}
 \frac{\tanh(\beta\epsilon_\kv/2)}{2\epsilon_\kv}.
 \label{SM:eq:curvature}
\end{equation}
Consequently, all considered single-channel instabilities occurs at
\begin{equation}
 J^\alpha<-\frac{U_c(L)}{4},\hspace*{0.3cm}\text{with}\hspace*{0.3cm} U_c(L)=\chi_0(L)^{-1}.
 \label{SM:eq:onset}
\end{equation}
This gives the reference onset lines used in the main text.
It uses the common dispersion and generator normalization.
Naturally, the onset lines do not need to coincide with phase boundaries accounting for interaction between competing FFs, as will be considered in the following section.

\section{Integration of FFs and their measures}
\label{SM:sec:marginals}

Accounting for interplay between FFs in MCFF theory is done through the free energies for the different FFs, where the remaining FFs are numerically integrating out.
When performing this integration, each FF integral has a weight $\exp[-\beta\mathcal F(\{\phi\})]$ and a flat measure in the retained Cartesian FFs.
Moreover, in the separate charge- and spin-sector calculations, the doublets with equal stiffness parameters can be described by:
\begin{equation}
 R_c=\sqrt{(\phi^{x,c})^2+(\phi^{z,c})^2},\hspace*{0.3cm}\text{and}\hspace*{0.3cm}
 R_s=\sqrt{(\phi^{x,z})^2+(\phi^{z,z})^2}.
 \label{SM:eq:radii}
\end{equation}
Since their angles do not enter their corresponding free energies, standard angular integration gives
\begin{equation}
 d\mu_c=2\pi R_c\,dR_c\,d\phi^{y,c},\hspace*{0.3cm}\text{and}\hspace*{0.3cm}
 d\mu_s=d\phi^{0,z}\,2\pi R_s\,dR_s\,d\phi^{y,z}.
 \label{SM:eq:measures}
\end{equation}
We highlight that orbital density wave (ODW) and spin-orbital density wave (SODW) labels in the phase diagrams of the main text refer to these $R_c$ and $R_s$, respectively.

In the integration performed within this work, the charge density wave (CDW) is treated separately in the charge-sector map, where its stable region is isolated from those of the orbital-dependent FFs.
Likewise, in the joint calculation, conducted near $J=0$, we retain the three components of $\boldsymbol\phi_c$, the three components of $\boldsymbol\phi_z$, and $\phi^{0,z}$, with their Cartesian product measure, as CDW stiffness parameter is positive there.
Furthermore, in the joint ensemble, the two orbital angles cannot be integrated independently as constant factors because their relative orientation affects Eq.~\eqref{SM:eq:seven}.

We now illustrate the construction of the free energies for a FF experiencing the remaining FFs.
Considering a FF $x$, its free energy is obtained by integrating over the remaining FFs:
\begin{equation}
 \beta\mathcal F_x(x)=-\ln\int d\mu_{\ne x}\,
 e^{-\beta\mathcal F(x,\phi_{\ne x})}.
 \label{SM:eq:marginal}
\end{equation}
For a radial coordinate, the angular variables are included in this integration, but the radial Jacobian of $x$ itself is excluded.
Separate charge and spin integrals are numerically evaluated by straightforward quadrature, while the joint seven-field integral is sampled by Monte Carlo sampling.
Both methods give the same free energies: for a radial distribution, $\beta\mathcal F_x(x)=-\ln[P_x(x)/\mathcal J_x(x)]$ up to a constant, with $\mathcal J_x\propto x$ for a doublet, and $\mathcal J_x\propto x^{D-1}$ for the total amplitude of $D$ Cartesian FFs.
Following the MCFF stability criterion~\cite{SM:PhysRevB.108.035143,SM:PhysRevB.108.205156}, a nonzero global minimum of $\mathcal F_x$ identifies a developed channel amplitude, and a nonzero secondary (local) minimum identifies the existence of metastable configurations within that channel.
We stress that single-component free energy profiles displayed in the spin-sector figure, only retain one Cartesian component and are integrate over the other three FFs; they need not have the same shape as the doublet profile.

These projected free energies account for both amplitude and orientation fluctuations.
For example, a three-component vector of fixed length with uniformly distributed orientations has a uniform distribution for each Cartesian component.
A flat single-component profile can therefore accompany a finite vector amplitude.
More generally, a minimum at zero in every channel free energy does not require the joint distribution to be concentrated at zero field.
This is why the main text also examines the free energy profile of the total amplitude $R_{\mathrm{tot}}=(\sum_\alpha(\phi^\alpha)^2)^{1/2}$.
This nonzero minimum is distinguishes in the main text as purple-colored regions, even though no individual channel is selected by their free energies.

\makeatletter
\begin{NAT@thebibliography}{1}

\bibitem{SM:PhysRevB.102.224423}
A.~N. Rubtsov, E.~A. Stepanov, and A.~I. Lichtenstein.
\newblock Collective magnetic fluctuations in hubbard plaquettes captured by
  fluctuating local field method.
\newblock {\em Phys. Rev. B}, 102:224423, 2020.

\bibitem{SM:PhysRevB.108.035143}
E.~Linn\'er, A.~I. Lichtenstein, S.~Biermann, and E.~A. Stepanov.
\newblock Multichannel fluctuating field approach to competing instabilities in
  interacting electronic systems.
\newblock {\em Phys. Rev. B}, 108:035143, 2023.

\bibitem{SM:PhysRevB.108.205156}
E.~Linn\'er, C.~Dutreix, S.~Biermann, and E.~A. Stepanov.
\newblock Coexistence of $s$-wave superconductivity and phase separation in the
  half-filled extended hubbard model with attractive interactions.
\newblock {\em Phys. Rev. B}, 108:205156, 2023.

\bibitem{SM:PhysRev.54.918}
R.~Peierls.
\newblock On a minimum property of the free energy.
\newblock {\em Phys. Rev.}, 54:918, 1938.

\bibitem{SM:Bogolyubov:1958zv}
N.~N. Bogolyubov.
\newblock A variation principle in the problem of many bodies.
\newblock {\em Dokl. Akad. Nauk SSSR}, 119:244, 1958.

\bibitem{SM:feynman1972}
R.~P. Feynman.
\newblock {\em Statistical Mechanics: A Set of Lectures}.
\newblock Benjamin/Cummings, Reading, MA, 1972.

\end{NAT@thebibliography}
\makeatother
\endgroup

\end{document}